\documentclass[letterpaper,twocolumn,10pt]{article}
\usepackage{usenix2019_v3}
\usepackage{tikz}
\usepackage{amsmath}

\usepackage{multirow}
\usepackage{subcaption}
\usepackage{graphicx}
\usepackage{arydshln}

\usepackage{amssymb}

\usepackage[ruled,vlined]{algorithm2e}

\newcommand{\basesol}{\texttt{\footnotesize baseSolution}}
\newcommand{\candsol}{\texttt{\footnotesize candidateSolution}}
\newcommand{\comsubsol}{\texttt{\footnotesize commonSub-Solution}}
\newcommand{\comsol}{\texttt{\footnotesize commonSolution}}
\newcommand{\comm}{\textbf{\texttt{\footnotesize comm}}}
\newcommand{\pref}[1]{\texttt{\footnotesize pref{#1}}}
\newcommand{\suff}[1]{\texttt{\footnotesize suff{#1}}}

\newcommand{\range}[2]{[#1, #2]}
\newcommand{\size}[1]{\big| #1 \big|}
\newcommand{\mydb}[1]{\langle #1 \rangle}

\iftrue
\newcommand{\dnote}[1]{[{\footnotesize \color{red}{\bf
dana:} { {#1}}}]}
\newcommand{\anote}[1]{[{\footnotesize \color{blue}{\bf
Aria:} { {#1}}}]}
\newcommand{\lnote}[1]{[{\footnotesize \color{orange}{\bf
Lambros:} { {#1}}}]}
\newcommand{\mnote}[1]{[{\footnotesize \color{green}{\bf
Mahammad:} { {#1}}}]}
\fi

\iftrue
\renewcommand{\dnote}[1]{}
\renewcommand{\anote}[1]{}
\renewcommand{\lnote}[1]{}
\renewcommand{\mnote}[1]{}
\fi

\usepackage{listings}
\definecolor{mGreen}{rgb}{0,0.6,0}
\definecolor{mGray}{rgb}{0.5,0.5,0.5}
\definecolor{mPurple}{rgb}{0.58,0,0.82}
\definecolor{backgroundColour}{rgb}{0.95,0.95,0.92}
\lstdefinestyle{CStyle}{
    backgroundcolor=\color{backgroundColour},   
    commentstyle=\color{mGreen},
    keywordstyle=\color{magenta},
    numberstyle=\tiny\color{mGray},
    stringstyle=\color{mGreen},
    basicstyle=\scriptsize,
    breakatwhitespace=false,         
    breaklines=true,                 
    captionpos=b,                    
    keepspaces=true,                 
    numbers=left,                    
    numbersep=5pt,                  
    showspaces=false,                
    showstringspaces=false,
    showtabs=false,                  
    tabsize=2,
    language=C
}

\begin{document}

\date{\today}

\title{\Large \bf Database Reconstruction from Noisy Volumes:\\
A Cache Side-Channel Attack on SQLite}


\author{
{\rm Aria Shahverdi}\\
University of Maryland\\
ariash@umd.edu
\and
{\rm Mahammad Shirinov\thanks{Part of this work was done while the author was an intern at the University of Maryland.}}\\
Bilkent University\\
shrnvm@gmail.com
 \and
 {\rm Dana Dachman-Soled\thanks{Work supported in part by NSF grants \#CNS-1933033,  \#CNS-1840893, \#CNS-1453045 (CAREER), by a research partnership award from Cisco and by financial assistance award 70NANB15H328 and 70NANB19H126 from the U.S. Department of Commerce, National Institute of Standards and Technology.}}\\
University of Maryland\\
danadach@umd.edu
} 

\maketitle

\begin{abstract}
We demonstrate the feasibility
of database reconstruction
under a
cache side-channel attack on SQLite.
Specifically, we present a Flush+Reload attack on SQLite
that obtains approximate (or ``noisy'') volumes of range queries made to a private database. We then present several algorithms that, taken together, reconstruct
nearly the exact database in varied experimental conditions, given these approximate volumes. 
Our reconstruction algorithms employ novel techniques for the approximate/noisy setting, including a noise-tolerant clique-finding algorithm,
a ``Match \& Extend'' algorithm for extrapolating volumes that are omitted from the clique,
and a ``Noise Reduction Step'' that 
makes use of the closest vector problem (CVP) solver to improve the overall accuracy of the 
reconstructed database.
The time complexity of our attacks grows quickly \dnote{exponentially?} with the size of the range of the queried attribute, but scales well to large databases. 
Experimental results show that we can reconstruct databases of size $100,000$ and ranges of size $12$ with an error percentage of $0.11 \%$ in under 12 hours on a personal laptop.
\dnote{Changed the above to ``depend mainly''. B/c when size of database grows, there could be a larger number of noisy volumes, resulting in more nodes and edges in the graph?}\anote{it is mostly true, although the number of nodes does not grow much}

\end{abstract}

\section{Introduction}

\dnote{Major things to do:
\begin{itemize}
    \item Section on "Assessing effectivenss of previous algorithms.."
    \item section on "Overview of Experimental Results."
    \item Experimental Section
    \item Conclusions section. We should probably say something about countermeasures there.
    \item Make sure we have included all the related work and given credit.
    \item Add claims (or revisit claims we already make) on run-time and/or uniform vs. non-uniform queries.
    \item Make sure all comments addressed
\end{itemize}
}
\anote{"Assessing effectivenss of previous algorithms.." and "Overview of Experimental Results." needs a pass}
\anote{Experimental Section }
\anote{run time of the algorithm grows exponentially with size of range but not with size of database, the NP-hard part regarding the clique}
\anote{conclusion is done}

Data processing in the cloud is becoming continually more pervasive and cloud computing is intrinsic to the business model of various popular services such as Microsoft's Office 365, Google's G suite, Adobe Creative Cloud or financial services such as intuit~\cite{cloudservices}.
Besides for cloud usage by industry, 
federal agencies are now utilizing cloud services,
even for storage and analytics of sensitive data.
For example, Microsoft recently won a $\$10$ billion government contract from the Department of Defense (DoD) to create a ``secure cloud'' for the Pentagon~\cite{DoDJedi}.
While providing important functionality, processing of sensitive information in the cloud raises important security challenges. In the extreme case, one may not trust the cloud server itself to handle the sensitive data, corresponding to a threat model in which the cloud server is assumed to be malicious. In this case, data must be encrypted, which raises the challenging task of computation over encrypted data. Techniques and tools for computation over encrypted data have been addressed in a myriad of papers~\cite{popa2011cryptdb, boldyreva2011order, lewi2016order, cash2013highly} and various privacy attacks have also been exhibited~\cite{naveed2015inference, kellaris2016generic}. 
A weaker threat model, considered in this work, assumes that the server may be trusted to handle the sensitive data (e.g. a privacy agreement has been signed with the cloud service), but that a spy process is running on the same public server. 
If a spy process is co-located with the victim on the same physical machine they will share hardware such as a cache, which serves as a side channel. 

Our goal is to explore the effect of side-channels on open-source database engines. We present an attack on SQLite, a C-language library that implements a small and fast SQL database engine and is among the top ten databases in the ranking released by db-engines.com.
Our threat model assumes that an external user queries a private database stored on a victim VM, upon which the victim VM processes the query using SQLite and returns the result to the external user. The attacker is disallowed from directly querying the database or observing the outputs of a query. Since the attacker is running a spy VM co-located with the victim VM in the cloud,
it can monitor the shared cache to obtain side-channel leakage. The goal of the attacker is to reconstruct the column upon which the victim is making range queries.


\paragraph{Relationship to attacks on Searchable Encryption.}
Our work is inspired by the line of works of
Kellaris et al.~\cite{kellaris2016generic},
Grubbs et al.~\cite{grubbs2018pump},
Lacharit\'{e} et al.~\cite{lacharite2018improved} and Grubbs et al.~\cite{grubbs2019learning}. These works exhibited database reconstruction attacks in scenarios where
range queries are made to an \emph{encrypted} database and
the access pattern (i.e.~which records are returned) \cite{kellaris2016generic, lacharite2018improved} or communication volume (i.e.~the number of records returned)~\cite{kellaris2016generic, grubbs2018pump} is observed by the malicious server. 
However, recall that in our threat model, an attacker cannot simply observe the access pattern or communication volume, 
and must instead resort to side channels 
(such as a shared cache) to learn information.
 Indeed, our attack will utilize the cache side-channel to learn
 information about the communication volume of the range queries.
 Briefly, this is done by finding a line of code that is executed once for each record returned in a response to a range query, 
 and tracking how many times that line of code is executed.
 
 Since cache side-channels are inherently noisy, we are only able to measure the \emph{approximate} or \emph{noisy} volumes of the range queries.
 We emphasize that even adding a small amount of noise to the volume of each range foils the
 reconstruction attacks from prior work.
 We assessed the effects of noise
 on brute force reconstruction (an analogue of the brute force algorithm suggested by~\cite{kellaris2016generic} for the dense database setting), and on the clique-finding approach developed by~\cite{grubbs2018pump}. 
 As will be discussed in depth in Section~\ref{sec:ourcontrib}\anote{here}, we conclude that both of these approaches fail in the noisy setting.

 \paragraph{Our approach.}
 We develop a new algorithmic approach that reduces our noisy problem to other computational problems that are well-studied in the literature and for which highly optimized solvers have been developed.
 Specifically, we will leverage both
 a noise-tolerant clique-finding algorithm (similar to \cite{grubbs2018pump}, but with some crucial modifications) as well as a closest vector problem (CVP) solver.
 In more detail, we first use the noisy cache data to craft an instance of the clique-finding problem that is noise-tolerant. Recovered cliques will then be used to obtain candidate databases that are ``close'' to the original database. To extrapolate volumes that may be entirely missing from the recovered cliques,
 we develop a ``Match \& Extend'' algorithm. After the Match \& Extend step,
 we expect to have reconstructed approximate volumes for all ranges.
 We then apply a ``Noise Reduction Step'' that takes the ``close'' solution outputted by 
 the previous step, consisting of approximate volumes for each
 of the ranges $\range{i}{i}$ for $i \in N$,
 and uses it to craft an instance of the CVP problem. Solutions to the CVP problem correspond to reconstructed databases in which the overall noise is further reduced.

We note that since our side-channel attack proceeds by measuring (approximate) range query volumes, 
it is agnostic to whether the victim's database is encrypted. As long as the spy can
monitor a line of code that is executed by the database engine for each record returned
by a range query, our attack is feasible.
Searchable encryption schemes that have this property would still be susceptible to this side-channel attack. 
For example searchable encryption schemes that can be integrated with standard database engines, such as order preserving encryption~\cite{SIG:AKSX04,EC:BCLO09} and order revealing encryption~\cite{EC:BL0SZZ15}.

 A limitation of our work is that our approach uses solvers for NP-hard problems as subroutines. The complexity of these NP-hard problems grows
 quickly \dnote{exponentially?}
 with the \dnote{square of the} size of the range, and therefore will work well in practice for ranges up to size 15. This is in contrast to the recent work of Grubbs et al.~\cite{grubbs2019learning}, which showed how to do ``approximate reconstruction'' in a way that scales only with the desired accuracy level and not the range size.
 However, the work of Grubbs et al.~\cite{grubbs2019learning} assumes the adversary gets to perfectly observe the \emph{access pattern}---i.e. which records are returned for each query---which provides far more information than simply observing the volumes. It seems difficult to extract the \emph{access pattern} for a
 response to a database query from a cache side-channel attack.\footnote{To extract access pattern from the cache, Prime \& Probe must be used to monitor the data cache, and because a single record from the database can fill a large portion of the cache, it is difficult to distinguish which records were accessed by observing only the cache. Additionally, the mapping from the memory location to cache line is not one to one and hence, a large number of records will map to the same cache locations, making it difficult to distinguish which records were accessed.
 }

We extensively test our attack in various scenarios, using real databases (with data distribution close to uniform), as well as synthetic databases with Gaussian data and various settings of the standard deviation. We also experiment with uniform queries (each possible range query is made with equal probability) and non-uniform queries (different range queries are made with different probabilities). We also extend our analysis to study the effect of extra load on the system. Furthermore, we extend the Match \& Extend algorithm by studying what will happen if not all the possible ranges are queried. 

\paragraph{Formal setting.} We consider a database of size $n$ and an attribute with range size $N$ for which range queries (i.e.~SQL queries that return all records corresponding to values between $[a,b]$) can be made. The size of query response corresponding to range query $\range{a}{b}$ is denoted by $\size{\range{a}{b}}$ and, similar to Grubbs et al.~\cite{grubbs2018pump}, the volumes in the form of $\range{1}{i}$ for $1 \leq i \leq N$ are called ``elementary volumes''. Note that to fully recover all the ranges in the form $\range{i}{i}$ it is enough to recover ``elementary volumes''. 
In this paper each volume is represented by a node in a graph. A node with label $v_i$ correspond to a range of volume $v_i$.
In this paper we refer to each node of the graph by its label.  The goal of our attack is to reconstruct the entire column corresponding to the field with range size $N$. Specifically, for each $i \in [N]$, we would like to recover the number of records $n_i$ that take value $i$ in the attribute under inspection. The focus of this work is on ``dense'' databases, meaning that every possible value from $1$ to $N$ is taken by some records in the database.\footnote{We note that in the searchable encryption setting this is not the typical case since ciphertexts encrypting values between $1$ and $N$ are typically sampled from a larger space. However, in this work, our main focus is on cleartext databases and attackers who learn information about them via the cache side-channel.} For simplicity, we assume that ranges are always from $1-N$. However, the result generalizes to any range $a-b$, where database records can take on at most $N$ discrete values within the range.
Our attack model assumes that a malicious party can only launch side-channel attacks to reconstruct the database. In particular, we assume that the attacker monitors its read timing from a cache line to deduce useful information about the victim. As discussed, the noise introduced by the cache side-channel makes our setting more challenging.\footnote{We note that Grubbs et al.~\cite{grubbs2018pump} mentioned a type of side-channel where an attacker intercepts the connection between user and server and counts the TLS packets in order to obtain volumes of range queries, but they did not consider the difficulties that arise when the measurement channel introduces noise into the computed volumes.}

\subsection{Our Contributions}\label{sec:ourcontrib}

We next summarize the main contributions of this work.

\textbf{Weaker threat model: Side-channels.} Prior work considers a threat model of a malicious server that is computing on an encrypted database. We consider an honest server computing on a cleartext (or encrypted) database and a malicious third-party that is co-located with the honest server in the cloud, sharing a cache, and cannot issue queries to the database. The malicious third-party can only obtain information by monitoring the shared cache. In particular, this means that the third-party cannot learn the \emph{exact} volumes of range queries and only obtains \emph{approximate} or \emph{noisy} volumes.

\textbf{Assessing effectiveness of
previous algorithms in the noisy setting.}
We first analyzed the effectiveness of a brute force attack,
similar to the one suggested in the work of Kellaris et al.~\cite{kellaris2016generic}, but adapted to the \emph{noisy}
and \emph{dense} database setting.
When we ran this version of the brute-force search algorithm, it failed to return a result, even after a day of running. 
We expected this to be the case, since when the volumes are noisy, there are far more choices that need to be checked in each step of the brute force search.

We next analyzed the effectiveness of an attack based on clique-finding, as in the work of Grubbs et al.~\cite{grubbs2018pump}.
A graph is constructed based on the observed volumes of the range queries.
To construct the graph from \emph{exact} volumes, one first creates nodes with labels corresponding to their volume, i.e. the node with label $v_i$ has volume $v_i$. There is a connection between node $v_i$ to $v_j$ if there exists a node $v_k$ such that $v_i = v_j + v_k$. Note that by this construction the nodes corresponding to elementary volumes form a clique of size $N$ which can be recovered by clique finding algorithm. 
The ranges $\range{1}{1}, \range{1}{2}, \ldots, \range{1}{N}$ and the full database can then be recovered from this information.

In the noiseless setting we always expect to get a clique of size $N$; however, in the noisy setting there are multiple edges missing
in the constructed graph and so a clique of size $N$ will typically not exist.
For example, when we ran the algorithm on our noisy data with $N = 12$, the size of the cliques returned was at most $3$. 
Further, the clique of size $3$ no longer corresponds to the volumes of the elementary volumes,
and therefore is not useful for (even partial) reconstruction of the database.


\textbf{Developing algorithms for the noisy setting.} 
Whereas Grubbs et al.~\cite{grubbs2018pump} used exact volumes to reduce the database reconstruction to a clique-finding problem, we begin by reducing the reconstruction problem given with volumes to a \textbf{Noise-Tolerant Clique Finding Problem} by introducing a notion called a \emph{noise budget}. 
Remember in the \emph{exact} volume case, there is a connection between node $v_i$ to $v_j$ if there exists a node $v_k$ such that $v_i = v_j + v_k$. 
Here, to construct the graph from \emph{noisy} volumes, we create a window, $w(v_k)$, of acceptable values around each leaked volume $v_k$, where the width of the windows is determined by the noise budget. We place an edge between node $v_i$ and $v_j$ if there exists a node $v_k$ such that $| v_i - v_j | \in w(v_k)$. 
The clique finding algorithm will return a clique that allows one to recover the volumes,
and the full database can then be recovered from this information. 
An attacker can determine a good setting of the \emph{noise budget} by mounting an attack in a preprocessing stage on a different, known database under same or similar conditions.
Specifically, the attacker can first create its own known database (unrelated to the unknown private database). The attacker can then simulate the side-channel attack on the known database on a similar system and compare the recovered approximate/noisy volumes with the correct volumes, and observe by how much they are off, to determine an appropriate \emph{noise budget}.
In some cases, incorporating the \emph{noise budget} into the construction of the graph and running the clique-finding algorithm already allows us to successfully reconstruct a fairly accurate database. However, there are some cases where, even after increasing the \emph{noise budget}, the algorithm fails to recover a candidate database (i.e.~a clique of size $N$ does not exist).
Further, even in cases where increasing the \emph{noise budget} does allow for reconstruction of some candidate database, the accuracy of the candidate database suffers and the run-time increases. We therefore introduce an additional algorithm called \textbf{Match \& Extend}, which allows successful reconstruction of candidate databases with improved accuracy.
 
 The \textbf{Match \& Extend} algorithm starts by obtaining a candidate clique from the graph. If the size of the clique is equal to $N$ (the maximum range) we are done. Otherwise, the algorithm looks at all the other cliques present in the graph starting from the largest to the smallest. For each clique, a potential database is recovered. We then pick one of the databases as our base solution and compare it with the other recovered databases. In the \textbf{Match} phase, the algorithm looks for the ``approximate longest common substring'' between two databases. The ``approximate'' version of the longest common substring considers two substrings equal if their corresponding values are within an acceptable range dictated by the \emph{noise budget}. Two values $a$ and $b$ are ``approximately'' equal if $ | a - b | \leq \min(a,b) \cdot 2 \cdot \texttt{\small noise budget}$. Then for the databases which have enough overlap with the base solution, the \textbf{Extend} phase will compare the non-matching parts of the two solutions and will try to reconcile the volumes in them into one ``combined'' database.
        
        
Finally, in the \textbf{Noise Reduction Step}, we use the results of the previous steps along with a closest vector problem (CVP) solver to reconstruct nearly the \emph{exact} original database, despite the noisy measurements. The recovered database of the previous step returns the ranges of the format $\range{1}{1}, \range{2}{2}, \ldots, \range{N}{N}$. We can reconstruct potential volumes for each range with these recovered volumes and for each computed volume we select the closest volume from the initial noisy volume set obtained from the side-channel data. We construct a lattice basis using the known linear dependencies between the volumes of different ranges. The volumes obtained from the side-channel data correspond to the target point for the CVP problem. Using the CVP solver, we find a set of volumes contained in the lattice (so they satisfy the linear dependencies) that are closest to the target point. This ``self-correction'' technique allows us to recover a better candidate solution for the database.

\textbf{Launching the side-channel attack.} 
We adapt the Flush+Reload technique for obtaining the (approximate) volumes of responses to range queries in SQLite. This allows us to learn a set of \emph{noisy} volumes corresponding to the range queries made by external parties to the database stored by the victim. The monitoring process starts as soon as an activity is detected and continues for the duration of the SQLite query processing. Since the databases we attack are large, the processing takes an extended amount of time, meaning that there are many opportunities for noise to be introduced into a trace. On the other hand, we require accurate measurements for our attack to succeed.\footnote{Similar to keystroke timing attacks of Gruss et al.~\cite{gruss2015cache}}
We contrast our setting to other side-channel settings, which typically require \emph{accurate} measurements over a \emph{short}
period or, can tolerate \emph{ inaccurate} measurements over a \emph{longer} period.
For example, side-channel attacks on cryptographic schemes require accurate information to reconstruct the high-entropy keys, but typically take a short period of time, since the keys themselves are short. On the other hand, side-channel attacks for profiling purposes typically monitor an application for longer periods of time, but can tolerate noise well since their goal is just to distinguish between several distinct scenarios.

To achieve high accuracy over a long period of time, we must handle interrupts as well as false positives and false negatives. For interrupts, we must mitigate their effects by detecting and dropping those traces in which an interrupt occurs. There can also be false positives as a result of CPU prefetching, which we show how to detect.
False negatives occur if the victim process accesses the monitored line of
code after the spy ``Reloads'' the line, and before the spy
``Flushes'' the line.
We do not directly detect false negatives, but instead show how to deal with them algorithmically.

\subsection{Overview of Experimental Results} 
We ran our attacks in five different experimental settings including uniform and non-uniform queries on real databases and synthetic databases which were sampled from Gaussian (Normal) distributions with different standard deviations as well as in two sets of experiments where the system is under heavy load and other cases where some of the range queries are missing. The databases all contained $100,000$ rows with $135$ attributes. The synthetic database from the Gaussian distribution has the same number of entries and attributes, but the column on which the range queries are made is sampled from Gaussians with standard deviation of $3$ and $4$, which represent narrow and wide Gaussians, respectively. The Match \& Extend algorithm recovered the database in $100 \%$ of the cases within $190$ seconds with maximum error percentage of $0.11 \%$.

\dnote{Just give the bottom line here...}\anote{fixed}

\anote{a paragraph here on the experimental result, final algorithm, our algorithm managed to recovered 100 percents to within 0.01 accuracy. Report the final Match \& Extend, maybe not the other attacks scenarios}

\dnote{Quick recap of when our attacks succeed.}
\dnote{Should discuss here what exactly are the requirements on the query distribution, and how our requirements compare to those of prior work.}


\subsection{Related Work}
\textit{Cache Attacks} were introduced by Tsunoo et al.~\cite{tsuno_cache} that shows a timing attack on MISTY1 block cipher.
Later, Osvik et al.~\cite{osvik2006cache} presented an attack that allowed the extraction of AES keys. In another early work, Ac\i i{\c{c}}mez~\cite{aciiccmez2007yet} showed an attack that targets instruction cache.
Ristenpart et al.~\cite{ristenpart2009hey} demonstrated the possibility of launching cache side-channel attacks in the cloud (as opposed to on a local machine) and they pointed out that such vulnerabilities leak information about the victim VM. Subsequent work showed how the cache side-channel can be used to extract cryptographic keys for ElGamal~\cite{zhang2012cross}, AES~\cite{irazoqui2014wait}, RSA~\cite{184415} and recently BLISS~\cite{bruinderink2016flush} (a lattice-based signature scheme).
In more recent work, Yarom and Falkner~\cite{184415} presented a powerful attack using Flush+Reload on the Level 3 cache of a modern processor. They tested their attack in two main scenarios, (a) victim and spy running on two unrelated processes in a single operating system and (b) victim and spy running on separate virtual machines. Another attack of note by Yarom and Benger~\cite{yarom2014recovering} on ECDSA leaks the nonce which results in signature forgery. 
A recent work by Moghimi et al.~\cite{moghimi2017cachezoom} showed the vulnerability of AES encryption in an SGX environment which, prior to this attack, was broadly believed to be secure. 
Ge et al.~\cite{ge2018survey} surveyed recent attacks and classified them according to which shared hardware device they target. Yan et al.~\cite{telepathy} shows the effectiveness of Flush+Reload and Prime \& Probe to reduce the search space of DNN architectures. In a more recent type of attack, Hong et al.~\cite{cacheNNFingerprint} shows how to perform Deep Neural Network fingerprinting by just observing the victim's cache behavior. 
In another work by Hong et al.~\cite{Hong2020How}, it is shown how to use cache attack to construct the main components of the Neural Network on the cloud.

\textit{Database Reconstruction} 
Kellaris et al.~\cite{kellaris2016generic}, motivated by practical implementations of searchable symmetric encryption or order-preserving encryption, studied the effect of auxiliary information on the overall security of the scheme. They identified two sources of leakage (a) access pattern (b) communication volume. They developed a reconstruction attack in which the server only needs to know the distribution of range query. They presented an attack using $N^4$ queries, where $N$ is the ranges of the value. Lacharit{\'e} et al.~\cite{lacharite2018improved} presents various types of attacks: full reconstruction, approximate reconstruction as well as a highly effective attack in which adversary has access to a distribution for the target dataset. Their attacks are based on the leakage of access pattern as well as leakage from the rank of an element. Grubbs et al.~\cite{grubbs2018pump} present an attack that reconstructs the database given the volumes of the response of range queries. They showed an attack using a graph-theoretic approach and specifically clique finding. Each volume is presented with a node in the graph. They demonstrated properties that hold in practice for typical databases and based on these properties they developed an algorithm which runs in multiple iterations of adding/deleting nodes. 
Once there is no more addition and deletion to be performed they announce that as the candidate database. They showed that this approach is indeed successful in recovering most of the columns of their example database. In cases where this algorithm could not find any possible result they used a clique algorithm to reconstruct the database, and they showed that clique could help to reconstruct even more instances.

In another line of work regarding searchable encryption, Cash et al.~\cite{cash2015leakage} presented leakage models for searchable encryption schemes and presented attacks. Specifically using this leakage they could recover queries as well as the plaintext. Naveed et al.~\cite{naveed2015inference} presented a series of attacks on Property-preserving Encrypted Databases. Their attack only used the encrypted column and used publicly known information. They showed an attack which could recover up to a certain attribute for up to $80\%$ of users. Grubbs et al.~\cite{grubbs2017leakage} presented an attack on order-preserving encryption and order-revealing encryption and showed they can reveal up to $99\%$ of encrypted values. Kornaropoulos et al.~\cite{kornaropoulos2019data} studied the database reconstruction given leakage from the $k$-nearest neighbours ($k$-NN) query. 
In a follow up work by the same authors, Kornaropoulos et al.~\cite{kornaropoulos2020state} extended their previous work by presenting an attack on encrypted database without the knowledge of the data or query distribution. All these attacks are in the encrypted database setting in which each value is encrypted whereas the focus of this work is on databases where the value of each entry is saved in clear text, and an attacker who may only obtain information about the database via side channels.
\dnote{are we missing important citations on attacks on searchable encryption?}

\section{Background} \label{sec:backgrounds}





\paragraph{Cache Architecture}
In order to reduce the access time to main memory, modern CPUs are equipped with multiple levels of cache. They form a hierarchy such that the Level 1 cache is the fastest and smallest, whereas the Level 3 cache is the slowest and largest. 

The Level 1 cache is divided into two separate caches, one holds the data and the other holds the instructions. In the higher level caches data and instructions are held in the same cache. Level 3 is a shared-memory space and is the Last Level Cache (LLC). The LLC is all-inclusive of the lower levels of the architecture, meaning that any data present in L1 and L2 is also present in the LLC.

Each cache comprises multiple sets and each set contains multiple cache lines. Each line of main memory is mapped to a unique cache set. Within this set, however, a memory line can be mapped to any of the cache lines. Typically, each line of cache holds 64 Bytes of data. Upon writing a line to a set that is already full, 
a decision regarding which memory line to evict must be made. This decision is called a \emph{Replacement Policy} and depends on the cache architecture. A popular replacement policy is \emph{least-recently used (LRU)}, which replaces the least recently used entry with the new one. 


\paragraph{Flush+Reload Attack}
Caches are vulnerable to information leakage since an adversary who is co-located with the victim on the same processor can retrieve useful information about a victim's activities.
Specifically, the adversary can monitor its own access time to the cache and use deviations in access time to deduce information about whether or not the victim has accessed a certain memory line or not. The reason that such an attack is feasible is that the adversary and victim share the same resource i.e.~the cache. 
Moreover, in a setting where the adversary and victim share a library, they will both have access to the physical memory locations in which the \emph{single copy} of the library is stored. The attacker can now explicitly remove a line corresponding to the shared physical memory from the cache. To exploit the shared physical memory in a useful way, Yarom and Falkner introduced an attack called \emph{Flush+Reload}~\cite{184415}.
The attacker flushes a monitored line from the cache using a special command called \texttt{clflush}. This command causes the monitored line to be removed from the L1, L2 and L3 caches. As mentioned before, L3 is inclusive and as a result the removed line will be removed from all the other caches, even if the attacker and the victim are not on the same physical core.
The attacker then lets the victim continue to run its program. After some time has elapsed, the attacker regains control and measures memory access time to determine whether or not the monitored line is present in the cache.
If the monitored line is present in the cache (reloading runs fast), the attacker deduces that the same line was accessed by the victim during its run. If the monitored line is not present in the cache (reloading runs slow), the attacker deduces that the victim did not access the line during its run. 
Hence the attacker knows whether the victim accessed a specific line or not. 
In order to perform the Flush+Reload attack we used the package provided in the Mastik framework. Mastik~\cite{yarom2016mastik} is a toolkit with various implementations of published micro-architectural side-channel attacks. It provides an interface that can be used to set the monitored lines. For our work we used \texttt{fr-trace} to monitor various cache lines. 



\paragraph{Cache Prefetching}
When an instruction or data is needed from memory, it is fetched and brought into the cache. To reduce execution time further, \emph{Cache Prefetching} is implemented to bring a memory line into the cache \emph{before} it is needed. The prefetching algorithm decides what and when to bring data and instruction to the cache. Hence, when the program needs the data or instruction in the future, it will be loaded from the cache instead of memory. This is based on the past access patterns or on the compiler's knowledge.

\paragraph{Range Queries}
A range query is an operation on a database 
in which records with column values between a certain lower and higher bound are returned. Assuming there exists a column $c$ in a database with values between $1$ and $N$, the command $\texttt{range}[a,b]$ for $1 \leq a \leq b \leq N$ returns all the entries in the database which have a value in column $c$ in the range $[a,b]$ (inclusive for both $a$ and $b$).

\paragraph{Clique Finding Problem}
The Clique problem is the problem of finding a clique--a set of fully connected nodes--in a graph. 
We utilize the clique finding algorithm in the NetworkX Package.\footnote{NetworkX is a Python library for studying graphs and networks.} The NetworkX package can be used to find the clique number (size of the largest clique in the graph) as well as all cliques of different sizes in the graph.

\section{Our Attack}\label{sec:attack}

In Section~\ref{sec:aprox_vol} we describe how to recover approximate volumes via the cache side-channel.
In Section~\ref{sec:clique} we describe how the clique-finding algorithm was used in the prior work of Grubbs et al.~\cite{grubbs2018pump} to recover a database from noiseless volumes. In Section~\ref{sec:noisy_clique} we explain our noise-tolerant clique-finding algorithm for our setting, where volumes are noisy. In Section~\ref{sec:match_extend} we present the details of the Match \& Extend algorithm which is used for extrapolating volumes that are omitted from the clique.
Finally in Section~\ref{sec:CVP} we describe how to use
closest vector problem (CVP) solvers to further reduce the noise and improve the overall accuracy of the recovered databases.


\subsection{Recovering Approximate Volumes}\label{sec:aprox_vol}


In this section, we explain how to find the lines of code in the SQLite library to monitor in the Flush+Reload attack and how to reduce the noise in our measurements. We will then explain how to recover the approximate volumes. 


\paragraph{Victim's Query} The victim issues a range query to SQLite database. SQLite returns the relevant entries as it processes the query. These entries are simply saved in a linked list and once SQLite is finished with processing the query, 
the linked list is returned to the victim.



\paragraph{Detecting Lines to Monitor}
SQLite stores columns using the BTree data structure.
We examined the SQLite program, and by using the \texttt{gcov} command we detected lines that are called once in each iteration of a range query. Monitoring the number of times these lines are called allowed us to determine the volume of a query response.
It is important to notice that the duration of each query can also be measured and that can also be used as an indicator for the volume. However, this resulted in far greater noise since there was no reliable way to translate time to volume (time to iterate over rows was inconsistent). Hence we decided to explicitly count.
To obtain the number of times each line is executed we compiled our library using \texttt{-fprofile-arcs} and \texttt{-ftest-coverage} flags.
We ran the range query command and by using the \texttt{gcov} command 
we counted the number of times each line is executed in files sqlite3.c and main.c, respectively. 

We looked for more lines throughout the SQLite\footnote{The attack presented in this paper can be extended to other database management system, as long as the volume of the returned query can be obtained through monitoring the I-cache.} program and we chose to simultaneously monitor \emph{two} lines to increase the measurement accuracy. 
As also observed by Allan et al.~\cite{allan2016amplifying}, monitoring two lines has the benefit that in case the attack code fails to detect an activity in one of the lines due to overlap between attacker reload and victim access \dnote{due to ***}\anote{addressed} there is still a high probability of seeing activity in the second line. There might be some excessive false positives due to the mismatch of hits for both of the lines and we mitigate for that by considering close hits to be from the same activity. 
\dnote{I remember that monitoring two lines can also cause false positives. Can you discuss how you detected/removed the false positives.}\anote{done}



\paragraph{Using the Mastik Toolkit.}
Once we detect the lines that leak the volume of the queries, we use the Mastik Toolkit to monitor those lines while SQLite is processing a range query.

\begin{figure}
 \centering
  \includegraphics[width=0.8\linewidth]{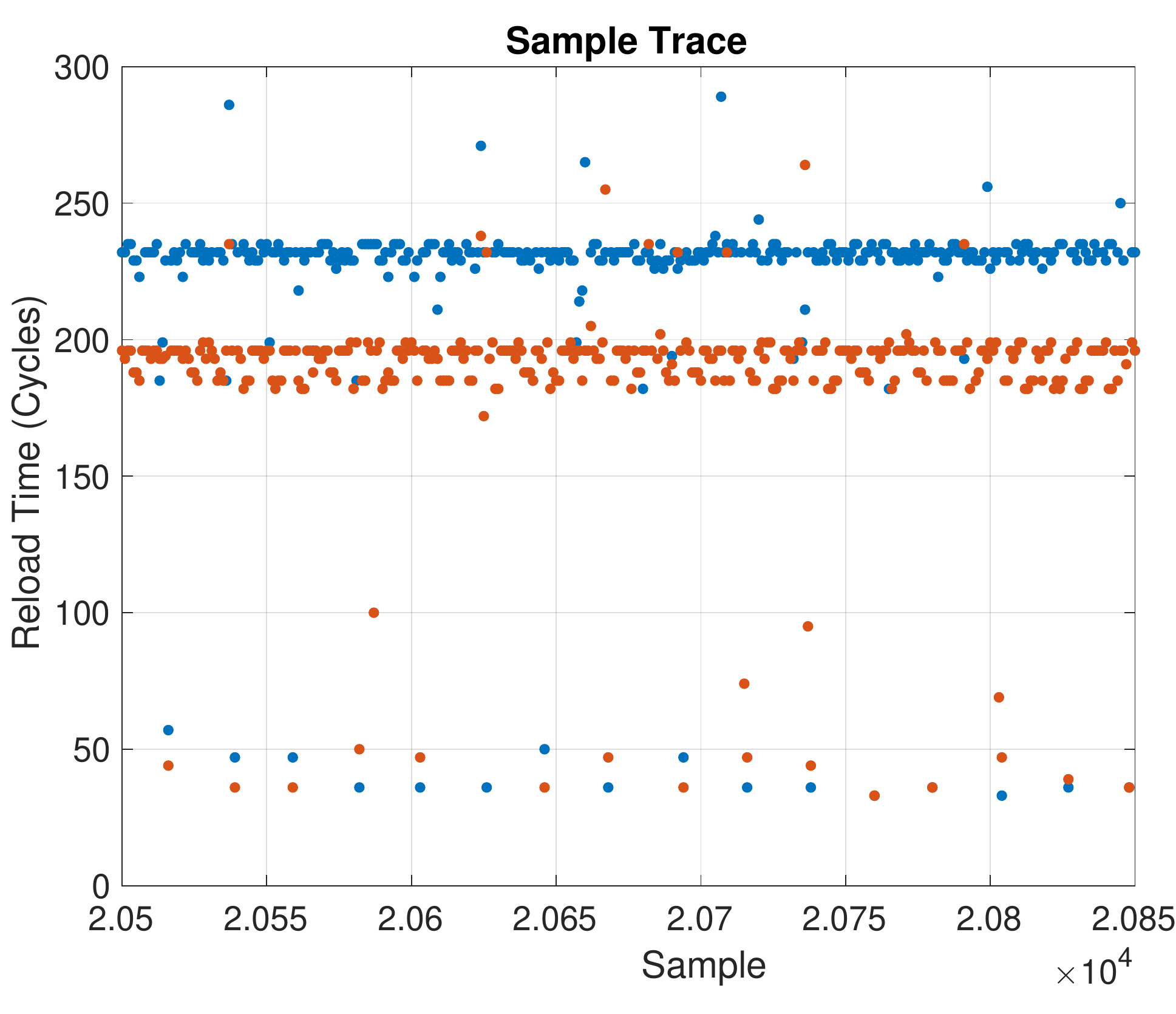}
  \caption{The result of running Flush+Reload attack on SQLite. There are two lines being monitored by attacker. The x-axis shows the sample point in which reload occurs and y-axis depicts the amount of time needed to reload the monitored line from memory at that time instance. Since two lines are being monitored, we have two sets of measurement at each time instance. Notice there are some orange lines appearing close to each other, those are because of speculative execution.}
  \label{fig:trace_details}
\end{figure}

Figure~\ref{fig:trace_details} shows one sample measurement. Two monitored lines are represented by blue and orange color. Mastik-\texttt{FR-trace} will automatically start measuring once it detects a hit in either of the monitored lines in the SQLite program. 
Once there is no more activity detected by Mastik for a while (as set in the \texttt{\small IDLE} flag), it will automatically end the measurement. During the interval where range query execution occurs, there are samples with reload time less than $100$ cycles. Those are the samples points in which SQLite accessed the line the attacker is monitoring and hence a small reload time is seen by the attacker. We then count the number of times there is a hit in either a blue or orange measurement. 
The hit count corresponds to the volume of the query. 

\begin{figure}[t]
 \centering
  \includegraphics[width=0.8\linewidth]{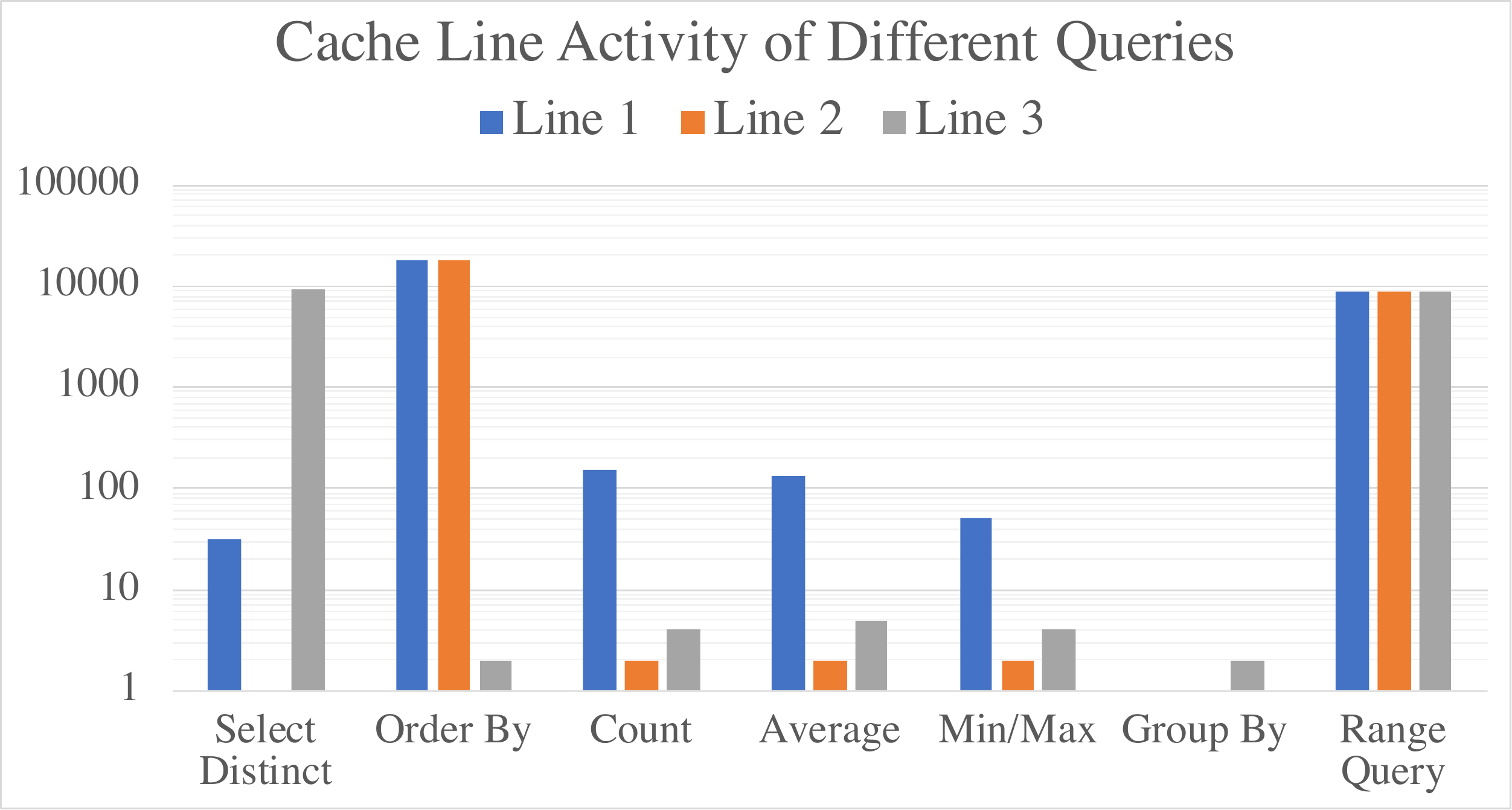}
  \caption{Cache line activity when different queries are issued. The cache line activity represents the number of hits detected by the attacker. The figure is in log scale. It can be seen that by looking at the cache activity of these three lines, different queries can be distinguished.}
  \label{fig:query_activity}
\end{figure}

Monitoring the relevant cache lines is also crucial to detect when/if range queries are issued. Figure~\ref{fig:query_activity} shows the cache activity of three cache lines in the first couple of thousands of samples. The cache line activity represents the number of hits detected by the attacker. Counting the number of hits represents whether or not the victim is using a specific line. We expect to see multiple cache activities in lines related to range queries and for the queries that are not relevant to range queries there is not much activity going on in at least one of the cache lines.

\paragraph{Noise in the traces}
The number of hits that we count might be different than the actual value of the volume since measurements are not noiseless. Here we explain some of the sources of noise. 
\begin{itemize}
    \item False Positive: Speculative execution of an instruction causes the memory line to be brought into the cache \emph{before} it is executed. In terms of the Flush+Reload attack, it will still \emph{look like} this instruction was executed, since there will be a fast access. Generally the true hits happen at fixed time intervals. If we see a hit which happens much sooner than the expected time for a hit it is most likely a false positive and we assume it occurred due to speculative execution and do not count it as a hit.
    \item False Negative: These occur if the victim process accesses the monitored line of code after the spy \emph{Reloads} the line, and before the spy
    \emph{Flushes} the line. 
    We do not attempt to detect false negatives experimentally, but rather deal with them algorithmically: as will be discussed in Section~\ref{sec:noisy_clique}, we use an asymmetric window around each observed volume to compensate for the fact that true volumes are typically greater than the observed volume. In our experiments we allocate $90 \%$ of the window width to the values greater than the observed volume. 
\end{itemize}



\paragraph{Running the experiment}
We randomly select 
and
execute range query $[a,b]$ while concurrently monitoring lines using Mastik-\texttt{FR-trace} to gather a single trace. We repeat this experiment a number of times in order to gather enough traces. For each trace we count the number of times that either of the lines shows a hit and after mitigating the False Positive issue, we report the number of hits as the volume of the range query for that trace.  

\begin{figure}[h]
 \centering
  \includegraphics[width=0.8\linewidth]{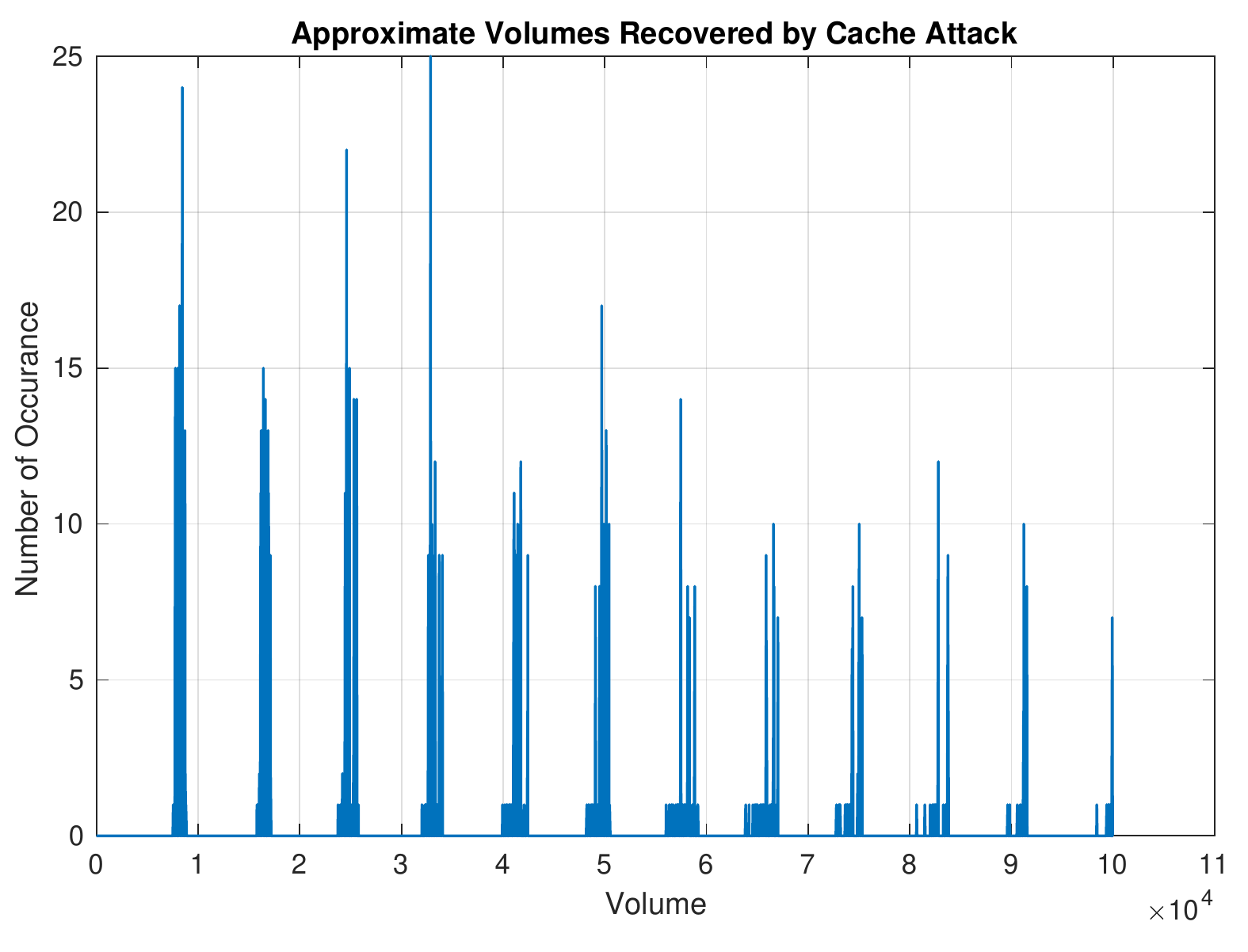}
  \caption{Sample noisy volumes recovered by cache attack. The x-axis is the volume and y-axis shows the number of occurrence of that volume. For a sample database, we ran the range query multiple times and for each range query we monitored the cache activity to recover the volume of the range for that query. We repeated this process multiple times and counted how many times a volume occurred.  
  }
  \label{fig:volume_attack}
\end{figure}

\begin{figure}
 \centering
  \includegraphics[width=0.8\linewidth]{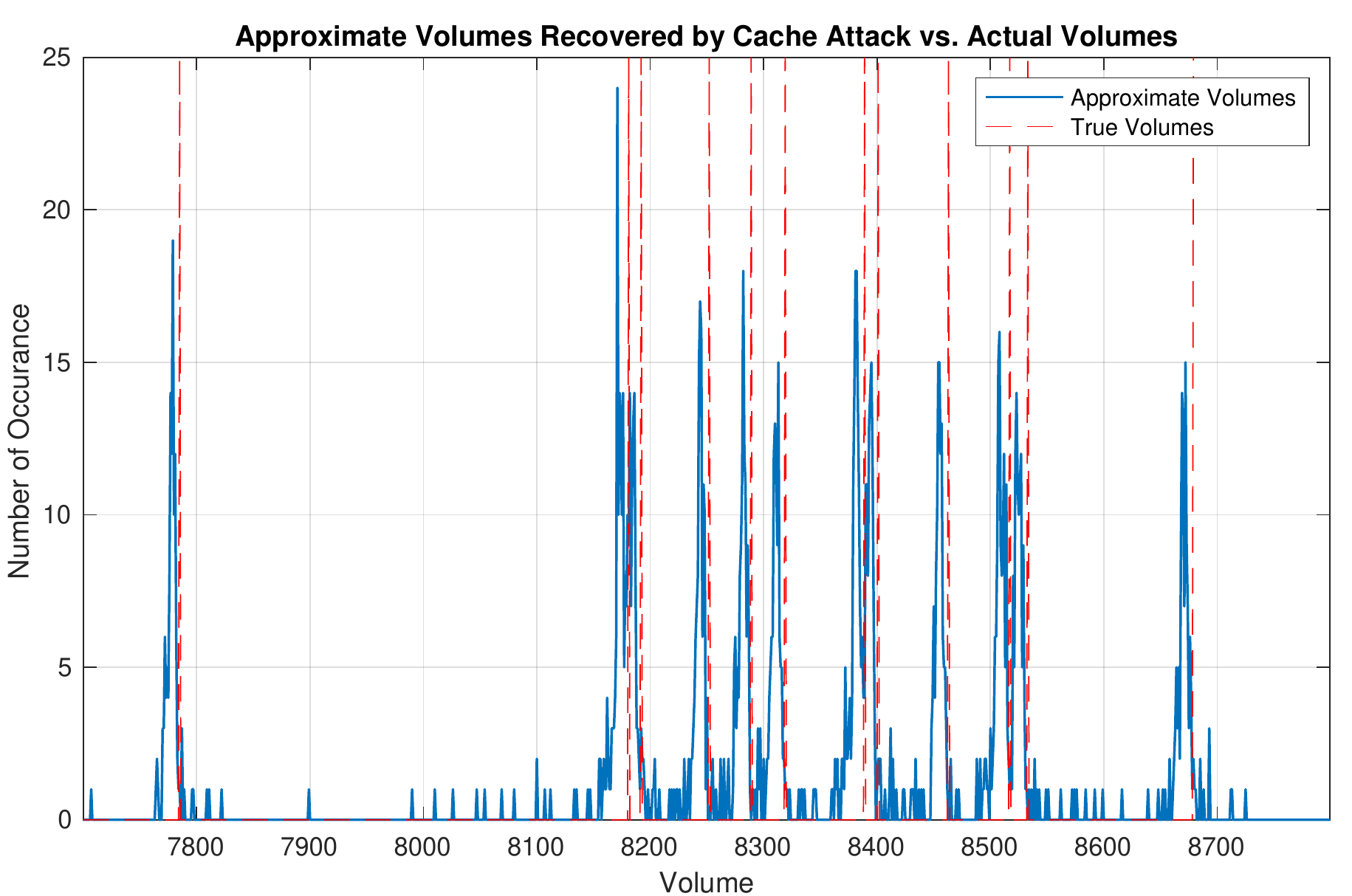}
  \caption{A closer look at the recovered volumes. The blue figure is the actual measurement from processing traces from the cache attack. The red bar is the actual volume expected to be observed. It can be seen that the recovered volumes (blue) is approximate version of actual volumes (red).}
  \label{fig:aprox_vs_real}
\end{figure}

Figure~\ref{fig:volume_attack} shows the result of aggregating the volumes reported by the traces. Some volumes are observed far more frequently than others, and those values are saved as an approximation to the expected volumes.
In a noiseless setting we expect to see at most $\binom{N}{2} + N$ values (there might be some volumes which correspond to more than one range query). In the noisy setting, there are cases where the trace is ``good enough'' but the volume is not correct. By aggregating all the traces the effect of those instances will be insignificant and the approximation of correct volumes will stand out. However, the volumes we recover are not exactly the correct volumes from the database.
Figure~\ref{fig:aprox_vs_real} shows a closer snapshot of Figure~\ref{fig:volume_attack} for volumes in the range $7700-8800$. The red dotted bars represent the actual volume of the range query response, while the blue line shows the approximate volumes recovered by the cache attack. For every correct volume (red line), there is a blue line with some high value close to it. 

\subsection{Clique Finding--Noiseless Volumes}\label{sec:clique}

To construct the graph we first explain the clique finding algorithm of Grubbs et. al.~\cite{grubbs2018pump} and then extend their technique to cover the noisy case. There are two main parts to the Algorithm. 
\begin{itemize}
    \item \textbf{Creating Nodes} Given the set of recovered volumes $V$ we create a node for representing each volume and label the node by its corresponding volume, meaning the node $v_i$ has volume $v_i$. 
    \item \textbf{Creating Edges} We create an undirected edge between two nodes $v_i, v_j \in V$ if there exists a node $v_k \in V$ such that $v_i = v_j + v_k$. 
\end{itemize}

By running the clique finding algorithm on the constructed graph, one can recover the volumes. Assuming the range of values are from $1$ to $N$ there are $\binom{N}{2} + N = \frac{N(N+1)}{2}$ possible ranges, and therefore $\frac{N(N+1)}{2}$ nodes in the graph. 
Each range $\range{i}{j}$ for $1 \leq  i \leq j \leq N$ is represented by a node. The nodes that correspond to ranges of the format $\range{1}{i}$ for $1 \leq i \leq N$, i.e. elementary volumes, form a clique, since for each pair of ranges of the form $\range{1}{i}$ and $\range{1}{j}$ for $1 \leq i < j \leq N$ there is another range of the form $\range{i+1}{j}$ 
for $1 \leq i < j \leq N$, which implies, due to how the graph is constructed, that there is an edge between 
$\range{1}{i}$ and $\range{1}{j}$.
The clique finding algorithm finds the nodes $\range{1}{i}$ for $1 \leq i \leq N$. To recover the original ranges which are of the form $\range{i}{i}$ for $1 \leq i \leq N$, all that is needed is to sort the nodes based on their labels, which corresponds to their volumes, and subtract them sequentially since $\size{\range{i}{i}} = \size{\range{1}{i}} - \size{\range{1}{i-1}}$ for $1 < i \leq N$. 

\subsection{Clique Finding--Noisy Volumes}\label{sec:noisy_clique}
In the noisy case considered here, all recovered volumes are close to the correct volumes, but the exact volumes may not have been recovered. 
Hence, the procedure for noiseless case fails to find the cliques of large enough size. This is because the condition to connect 
nodes $v_i, v_j$ will almost always fail
(even when there should be an edge)
since there will not be a third volume $v_k$ such that the equation $v_i = v_j + v_k$ is exactly satisfied. 
This means that the constructed graph is missing too many edges and the large cliques are not formed. 
To mitigate the effect of the noise, we modify the second step of the graph generation algorithm i.e. \textit{Creating Edges}. 


While the recovered volumes are close to the correct ones, as explained in Section~\ref{sec:aprox_vol}, since the traces are noisy we do not expect to get the exact volumes and we often under-count.
We call the ratio of the recovered volume to the correct volume the ``noise ratio''. In the first step the attacker performs a preprocessing step which involves mounting the attack on a database known to the attacker. The attacker then assesses the quality of the traces to find the approximate value of the ``noise ratio''. To find it the attacker heuristically looks at the recovered volumes and compares them to the correct volumes they are expecting to compute. Then based on all the noise ratios, the attacker sets a value for ``noise budget'' which is the mean of the 
``noise ratio'' he observed over all volumes. Once the noise budget is fixed, for each recovered volume the attacker creates a window of acceptable values around it. Assuming the recovered volume is $v_i$, the attacker creates an asymmetric window around $v_i$ with lower bound and upper bound of $v_i \times (1 - 0.1 \cdot \texttt{\small noise budget})$ and $v_i \times (1 + 0.9 \cdot \texttt{\small noise budget})$, respectively. As also mentioned in Section~\ref{sec:aprox_vol}, the window is asymmetric with $90\% $ of its width on the right hand side of it, as the noisy volumes are typically less than the true volumes.
For a volume $v_i$ we denote by $w(v_i)$ the window around it. To construct the graph in the noisy case we modify the second step of the algorithm explained in Section~\ref{sec:clique} as follows: 

\begin{itemize} 
    \item \textbf{(Modified) Creating Edges} We create an undirected edge between two nodes $v_i \in V$ and $v_j \in V$ if there exists a node $v_k \in V$ such that $| v_i - v_j | \in w(v_k)$. 
\end{itemize} 

In particular, we will say that candidate volumes $u$ and $v$ are ``approximately equal'' if $\frac{|u-v|}{min(u,v)} \le \texttt{\small noise budget}$. As we will show in Section~\ref{sec:experiments}, using the above algorithm with the just-mentioned modification 
is in some cases sufficient to approximately reconstruct the database.

\subsection{Match \& Extend}\label{sec:match_extend}

In this section we describe an improvement on the noisy clique-finding algorithm that is used in cases where the noisy clique-finding algorithm fails to find a maximal clique of size $N$, even with appropriate adjustment of the noise budget. 

\begin{figure*}[h]
\centering
\begin{subfigure}[t]{0.20\textwidth}
\includegraphics[width=\textwidth]{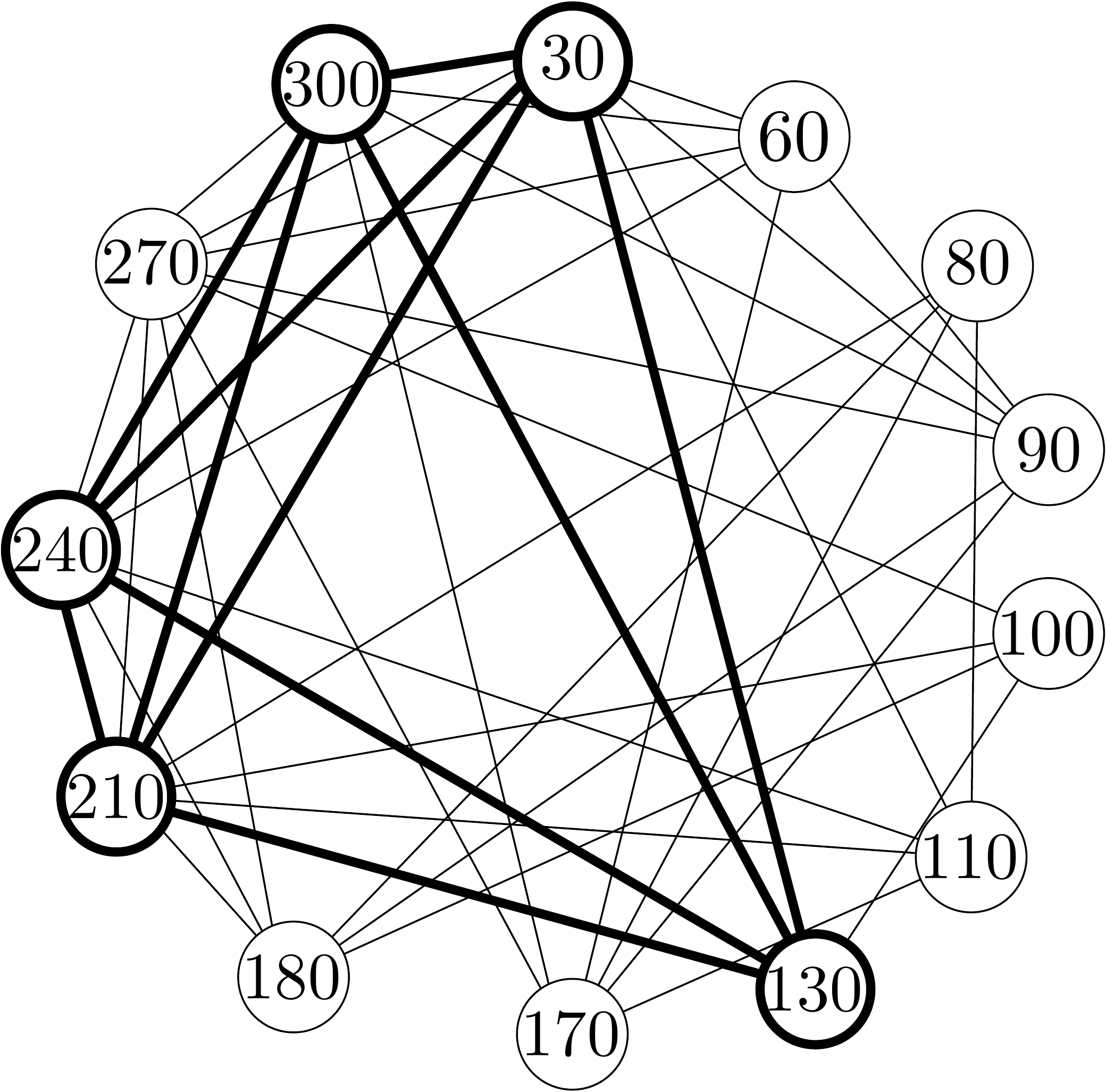}
\subcaption{The graph constructed from the \emph{exact} volumes of the database $[30, 100, 80, 30, 60]$ and the maximal clique corresponding to that.}
\label{fig:clique1}
\end{subfigure}
\hfill
\begin{subfigure}[t]{0.20\textwidth}
\includegraphics[width=\textwidth]{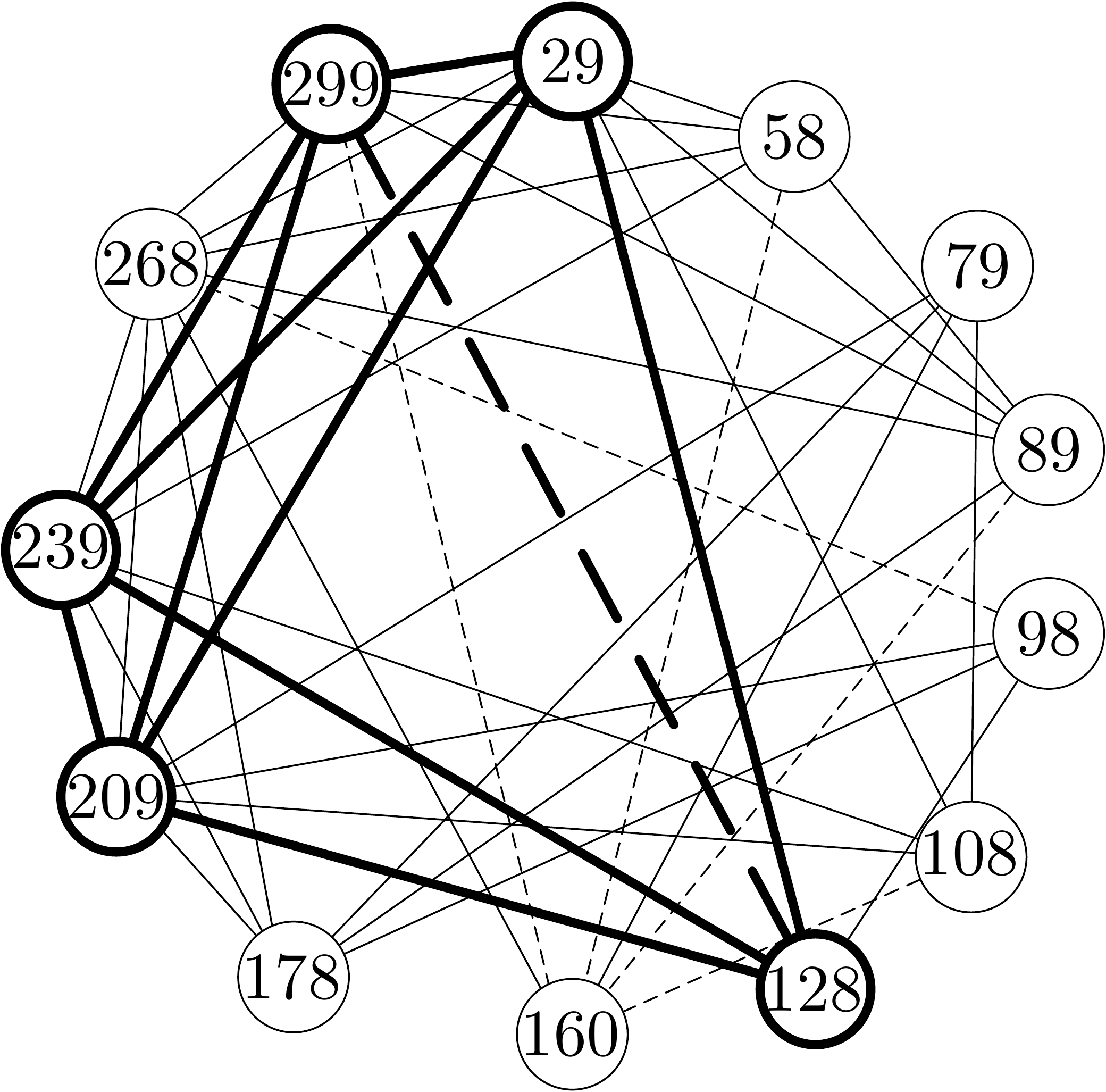}
\subcaption{The graph constructed from the \emph{approximate/noisy} volumes of the database $[30, 100, 80, 30, 60]$. An edge in the maximal clique is missing.}
\label{fig:clique2}
\end{subfigure}
\hfill
\begin{subfigure}[t]{0.20\textwidth}
\includegraphics[width=\textwidth]{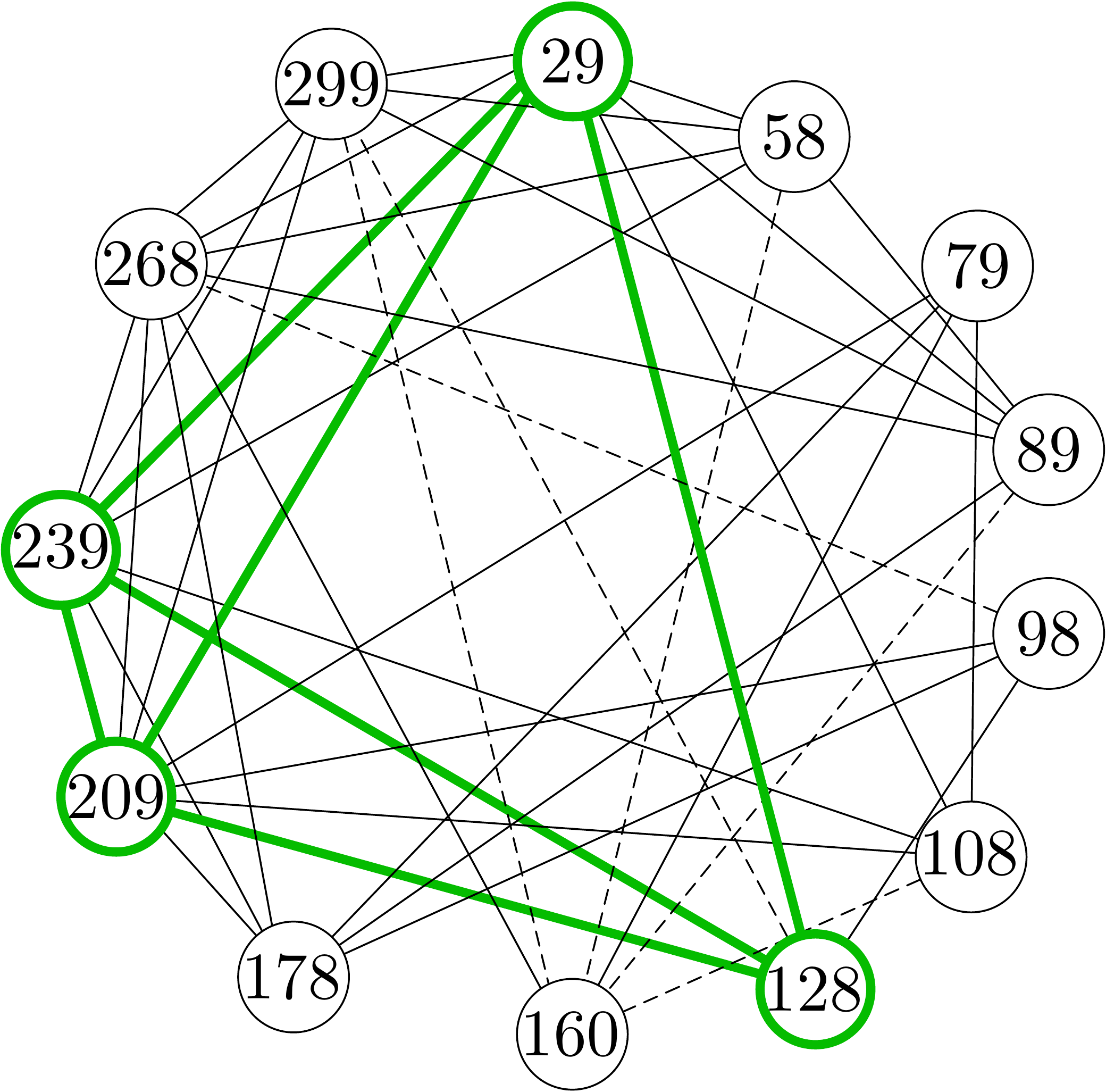}
\caption{A maximal Clique in a graph with \emph{approximate/noisy} volumes.}
\label{fig:clique3}
\end{subfigure}
\hfill
\begin{subfigure}[t]{0.20\textwidth}
\includegraphics[width=\textwidth]{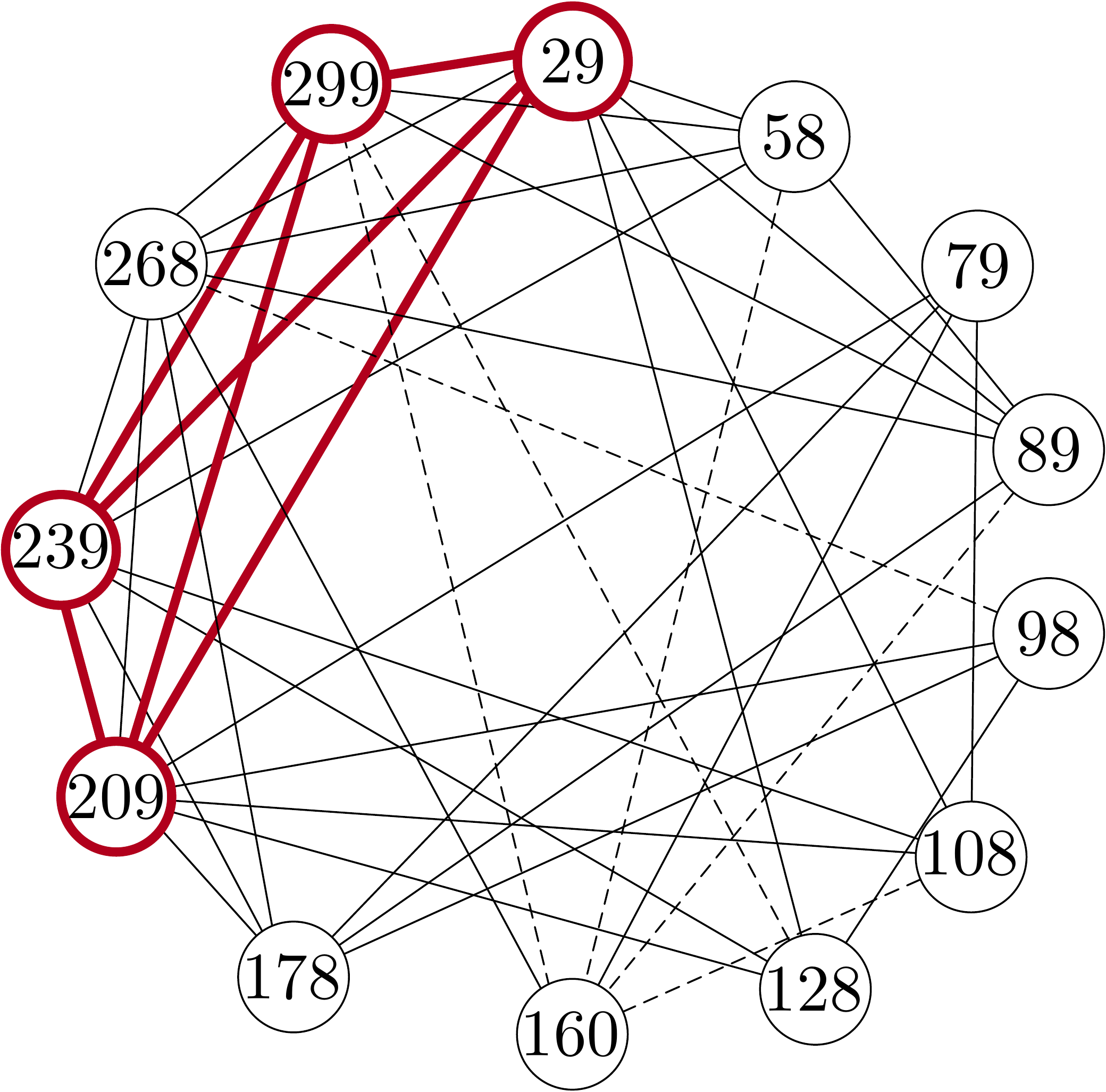}
\caption{Another maximal Clique in a graph with \emph{approximate/noisy} volumes.}
\label{fig:clique4}
\end{subfigure}
\caption{The graph constructed from the \emph{exact} volumes has a maximal clique which can be used to recover the original database, while the noisy volumes might have smaller maximal cliques and the original database can not be recovered from just one clique.}
\label{fig:explain}
\end{figure*}

First, recall that the idea behind the clique finding algorithm is that if we have the volumes of all ranges present in our data, then there must exist a clique in the graph corresponding to the volumes of the ranges $\range{1}{i}$ for $1 \leq i \leq N$. Now, let us assume there is a missing (approximate) volume corresponding to range $\range{i}{j}$.  This will result in the missing connection from the node $\range{1}{j}$ to node $\range{1}{i-1}$ as the reason that there had to be a connection was because $\size{\range{1}{j}} \approx \size{\range{1}{i-1}} + \size{\range{i}{j}}$. As a result of this missing volume, the maximal clique of size $N$ will not form. If we run the clique finding algorithm on the data with the missing volume, it will return cliques of size smaller than $N$ and for each of them recover a candidate database. Then the algorithm will merge the information in these smaller databases to form larger ones. In the following we explain the idea of the algorithm with an example.
Consider a database with $5$ possible values in the range, i.e. $N = 5$, assume the database is $\mydb{30, 100, 80, 30, 60}$ (i.e.~the database contains $30$ records with value $1$, $100$ records with value $2$, etc.). The set of possible values for the volume of a range query is $V = \{30, 60, 80, 90, 100, 110, 130, 170, 180, 210, 240, 270, 300\}$, i.e. $\size{\range{1}{1}} = 30$, $\size{\range{1}{2}} = 130$ and so on. The graph constructed from the these volumes is shown in Figure~\ref{fig:clique1} and the maximal clique found by the clique finder algorithm is shown by bold connections.
The returned nodes are $\{30, 130, 210, 240, 300\}$ and the reconstructed database is $\mydb{30, 100, 80, 30, 60}$.
Assume the recovered volumes are noisy and the set of possible values for the volume of a range query is $V = \{29, 58, 79, 89, 98, 108, 128, 160, 178, 209, 239, 268, 299\}$. 
In Figure~\ref{fig:clique2} all the noisy volumes are rather close to their actual values except the volume $160$ which is far from the correct one --- $170$. 
To construct the graph in this setting we use the algorithm mentioned in Section~\ref{sec:noisy_clique} and only for the sake of this example we take the window around volume $v_i$ to have lower and upper bound of $v_i - 1$ and $v_i + 3$, respectively. 
Some connections will be missing as a result of the error in the measurement. For example the connection from node $299$ to node $128$ is not going to be formed since there is no longer a window which contains $171$. If we run the clique finding algorithm on the new graph the result is going to be a clique of size smaller than $N = 5$. As seen in Figure~\ref{fig:clique3}, the clique finding algorithm returns a clique of size $4$ with values $\{29, 128, 209, 239 \}$ that results in database $\mydb{29, 99, 81, 30}$. Figure~\ref{fig:clique4} shows another clique of size $4$ with values $\{29, 209, 239, 299\}$ that results in database $\mydb{29, 180, 30, 60}$. It can be observed the two databases \emph{approximately} ``match'' in some locations, i.e. $\mydb{29, 180 (99 + 81), 30}$. It is important to note that although in this example $180$ is exactly equal to $99+81$, this need not hold in general, thus we consider two sequences a match if their corresponding values are \emph{approximately} equal. Having established this long match, we can deduce that value $60$ also belongs to the database and we can ``extend'' the initial candidate to include $60$ and return the database $\mydb{29, 99, 81, 30, 60}$. In another scenario assume we first detect the database of Figure~\ref{fig:clique4} and then we discover the database in Figure~\ref{fig:clique3}. In that case we can see that we can rewrite the initial candidate, i.e. $\mydb{29, 180, 30, 60}$ as $\mydb{29,180=(99+81),30,60}$ using the second candidate.

We next describe in detail the main steps taken in the Match \& Extend algorithm.
The high level steps of the Match \& Extend can be found in Algorithm~\ref{alg:match_extend}.
We note that, in some cases,
simply increasing the noise budget allows us to  successfully find a clique of size $N$. However, as we will discuss in Section~\ref{sec:experiments}, by not increasing the noise budget and instead running the Match \& Extend algorithm, we can recover a database that is closer to the true database.

\begin{algorithm}
\scriptsize
\SetAlgoLined
\KwResult{A database with $N$ values}
 \basesol\ = FindMaximalClique()\;
 allCliques = FindRemainingCliques(K, $\ell$)\;
 \While{length(\basesol) < N}{
 \candsol = FindBestCandidate(allCliques)\;
 \basesol = Merge(\basesol, \candsol)
 }
 \Return \basesol
 \caption{Match \& Extend Algorithm}
 \label{alg:match_extend}
\end{algorithm}


\paragraph{FindMaximalClique} The first step in the Match \& Extend algorithm is to find a maximal clique in the constructed graph. Let $K$ denote the size of a maximal clique recovered in this step. If there is more than one clique with the same maximal size select one of them arbitrarily. Once the clique is found, the corresponding database is computed. We call this database \basesol\ and the rest of the algorithm will expand this database. If the size of maximal clique found in this step is $N$ we are done, otherwise the Match \& Extend algorithm expands the \basesol.


\paragraph{FindRemainingCliques} Recover all cliques of size ${K, K - 1, K - 2, \ldots, K - \ell}$ and sort them from the largest clique size to the smallest. For each clique, the corresponding database is found and is called \candsol. The \candsol\ is in the form of an ordered list of volumes that correspond to neighbouring ranges of the database. Note that the cliques to be found in this step are not restricted to be from the ranges in the form $\range{1}{1}, \range{1}{2}, ..., \range{1}{K}$ for some $K$. In fact, and this holds in the noiseless setting too, any set of volumes corresponding to ranges of the form $\range{i}{i_1}, \range{i}{i_2}, \dots, \range{i}{i_k}$ where $i\leq i_1 < i_2 < \dots < i_k$ will form a clique of size $k$, provided that all the differences of the volumes corresponding to these ranges are present in our data. This fact will enable our algorithm to discover the volumes of different parts of the true database and ``merge'' those parts to recover the original database.


\paragraph{ApproximateLCSubstring} This is a subroutine that is invoked as a part of \textbf{Merge} function. Given a \basesol\ and a \candsol\ in form of lists of volumes of neighboring ranges, it finds the longest common substring of these solutions, i.e. the longest contiguous list of volumes where both solutions agree. We call this substring the \comsubsol. To find it, we use a standard \emph{longest common substring} algorithm with a modification that the elements of the substring need to be only approximately equal (as defined in Section~\ref{sec:noisy_clique}) to the corresponding elements of \basesol\ and \candsol. At termination this will return the \comsubsol\, and the starting and ending indices of \comsubsol\ in the two given solutions.


\paragraph{Merge} Given the \basesol\ and a \candsol\ in the form of lists of volumes, attempt to combine the information in them into one larger solution. We refer to this as ``merging'' the two solutions. The \textbf{Merge} function first invokes \textbf{ApproximateLCSubstring} to find the approximate longest common substring of the two solutions. After the longest common substring of the two solutions and the locations of this substring in the two solutions are found, there can still be volumes where the \basesol\ and \candsol\ agree, which are not recognized by the \textbf{ApproximateLCSubstring}. 
For example, if the \basesol\ is $\mydb{\textbf{29}, 99, 81, 30}$ and \candsol\ is $\mydb{\textbf{29}, 180, 30, 60}$, the \comsubsol\ may be found as $\mydb{\textbf{29}}$, however one can see that the two solutions agree at $\mydb{99, 81}$ as well, only in the \candsol\ this information appears as the volume of one range $\mydb{180}$ which is the union of those two neighboring ranges in \basesol. 
The merging algorithm identifies such cases and extends the \comsubsol\ accordingly. The algorithm searches for occurrences where a volume $v_i$ next to the end of the \comsubsol\ in one of the solutions (say in \basesol) is approximately equal to the sum of volumes $u_j, u_{j+1}, \dots , u_{j+r}$ for $r \ge 0$ next to the same end of the \comsubsol\ in the other solution (say \candsol). In such a case, it extends the \comsubsol\ by appending to it $\mydb{u_j, u_{j+1}, \dots , u_{j+r}}$, and changing endpoints of the \comsubsol\ in \basesol\ and \candsol. 
So in the database given above, the algorithm will look at the neighbors of $\mydb{\textbf{29}}$ and discover that $180 \approx 99 + 81$, and extend the \comsubsol\ to $\mydb{\textbf{29}, \textbf{99}, \textbf{81}}$. Then, the algorithm will look at the neighbors of $\mydb{\textbf{29}, \textbf{99}, \textbf{81}}$ and discover that $30 \approx 30$, extending the \comsol\ further to $\mydb{\textbf{29}, \textbf{99}, \textbf{81}, \textbf{30}}$. It is important to mention that while the values in the example were exactly equal, the algorithm accepts values which are approximately equal as well, meaning that we look for whether $180 \stackrel{?}{\approx} 99 + 81$ or whether $30 \stackrel{?}{\approx} 30$. 
After the \comsubsol\ is maximally extended, our two solutions will have the following form: $\basesol\ = 
\mydb{\pref{1}, \comm, \suff{1}}$ and $\candsol\ = \mydb{\pref{2}, \comm, \suff{2}}$, where \comm\ is the \comsubsol\ found as previously explained, and any of the prefixes and suffixes may be empty. 
The algorithm then will do one of four things: (a) if \pref{1} (similarly, \suff{1}) is empty, it will extend the \comsubsol\ to $\comm\ = \pref{2} || \comm\ $ (similarly, $\comm\ || \suff{2}$), (b) if \pref{2} (similarly, \suff{2}) is empty, it will extend the \comsubsol\ to $\comm\ = \pref{1} || \comm\ $ (similarly, $\comm\ || \suff{1}$), (c) if both \pref{1} and \pref{2} (similarly, \suff{1} and \suff{2}) are of length 1, meaning they both contain one volume (say, $a$ and $b$, with $a<b$), and if the absolute value of the difference of these volumes appears in our volume measurements, then $\comm\ = \mydb{b-a, a} || \comm\ $ (similarly $\comm\ = \comm\ || \mydb{a, b-a})$, (d) if none of the above conditions are satisfied, the algorithm will abort the merge and repeat its steps for another \candsol. The condition (c) above is for identifying the cases where the volume in, say, \suff{1} corresponds to a range $\range{i}{j}$, which includes in itself the range of the volume in \suff{2}, which can be $\range{i}{k}$ for $k<j$. If the difference of these two volumes appears in the measured volumes, that difference likely corresponds to the range $\range{k+1}{j}$, so we replace $\mydb{b}$ ($\range{i}{j}$) by $\mydb{a,b-a}$ (which is the range $\range{i}{k}$ and range $\range{k+1}{j}$, respectively).

Back to our example database of $\mydb{30, 100, 80, 30, 60}$, we had last found the \comsubsol\ to be $\mydb{\textbf{29, 99, 81, 30}}$. In the merge step, we are going to have $\basesol = \mydb{ \comm }$ and $\candsol = \mydb{ \comm, 60}$. This falls under the case (a) where \suff{1} is empty, so the algorithm appends \suff{2} = $\mydb{60}$ to the \comsubsol\ and returns the solution as $\basesol = \mydb{\textbf{29, 99, 81, 30, 60}}$.


\paragraph{FindBestCandidate} Any time a merge is successful, two solutions are combined into one to create a larger solution. The reason why this larger solution was not initially found by the clique finder is that some volumes or connections in the graph were missing, and so a potential clique corresponding to this solution could not be formed. Every merge of two solutions identifies the number of missing volumes that prevented the combined solution from being found in the first place; in fact, if we were to add those missing volumes to the graph and start the algorithm again, the combined ``merged'' solution would show up among all listed solutions. Therefore we use the number of missing volumes as a metric for assessing the \emph{goodness} of a candidate solution; if there are few missing volumes, it suggests that the \basesol\ and \candsol\ agree in many volumes of the database, and are thus compatible, whereas if there are many missing volumes, the two solutions likely have different information about the volumes. The \textbf{FindBestCandidate} finds the candidate solution among all cliques that has the least number of such missing volumes with respect to being merged with the \basesol.

\section{Experimental Results}\label{sec:experiments}

We performed five sets of experiments. The first two experiments (I and II) are for the cases where there is no additional noise during the measurement and the query distribution over all the possible queries are uniform. The third set of experiments (III) in Section~\ref{subsec:extraload} studies the effect of extra load on the system while taking measurements. The fourth set of experiments (IV) in Section~\ref{subsec:nonuni} looks at different query distributions. In the last set of experiments (V) in Section~\ref{subsec:missingvol}, we look at the effect of missing some volumes due to the fact that some range queries may have never been issued, or due to noise causing a query to be missed entirely. 
In Experiment I, we first prepare 10 databases from the  NIS2008 database, by randomly selecting $100,000$ records. 
Nationwide Inpatient Sample (NIS) is part of the Healthcare Cost and Utilization Project (HCUP) which is used to analyze national trends in healthcare~\cite{nis2880}. The NIS is collected annually and it gathers approximately 5 to 8 million records of inpatient stays. We selected NIS from the year 2008; the full description of each attribute of the database is reported in~\cite[Table1]{nis2880}. 
In the first set of experiments we performed uniform range queries on the AMONTH attribute which corresponds to admission month coded from (1) January to (12) December (i.e., each of the possible ranges were queried with equal probability). 
In the second set of experiments (Experiment II) we sampled the database as follows: For each of the $10$ databases from Experiment I, for each record in the database, instead of using the real value for the AMONTH column, we generated synthetic data by sampling a value from a Gaussian (Normal) distribution with mean $\frac{1+N}{2} = 6.5$ and standard deviation of $3$ and $4$, respectively. So within Experiment II, we considered two data distributions, a ``narrow'' Gaussian with standard deviation $3$ and a ``wide'' Gaussian with standard deviation $4$. 


We ran the experiments on a Lenovo W540 Laptop with Intel Core i7-4600M CPU clocking at $2.9$ GHz running Ubuntu 16.04. The L1, L2 and L3 caches have capacities 32KB, 256KB and 4MB, respectively. For the SQLite, we use the amalgamation of SQLite in C, version 3.20.1.
We heuristically observed that if we gather around $120$ measurements for any one range query, the aggregated side-channel measurements will result in a peak corresponding to the approximate volume.
Since there are at most $78$ different range queries for $N=12$ we decided to gather around $10,000$ traces to make sure there are enough traces for each range to be able to see a peak for each approximate volume.

We gathered $10,000$ traces corresponding to $10,000$ uniformly chosen range queries for Experiments I and II, i.e. $1$ trace for each query. We processed all those traces to obtain the approximate/noisy volumes. On average, gathering $10,000$ traces takes around $8$ hours and processing them takes another $3$ hours. The experiments and the code to run the Clique-finding algorithm, Match \& Extend and noise reduction step can be found here~\cite{github:repo}.

\begin{figure*}[h]
\centering
\begin{subfigure}[t]{0.3\textwidth}
\includegraphics[width=\textwidth]{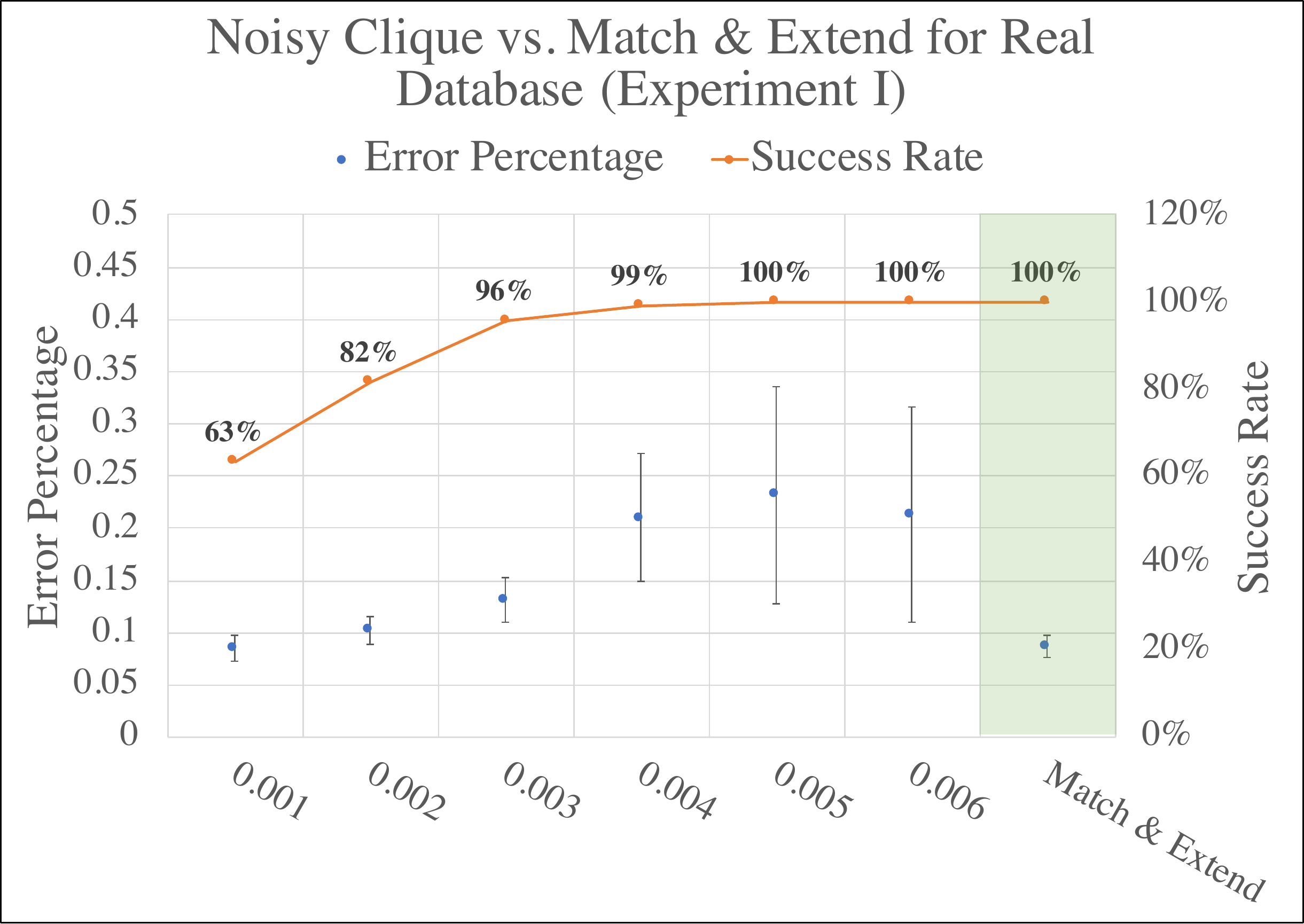}
\subcaption{Success Rate and error percentage for Experiment I}
\label{fig:exp1}
\end{subfigure}
\hfill
\begin{subfigure}[t]{0.3\textwidth}
\includegraphics[width=\textwidth]{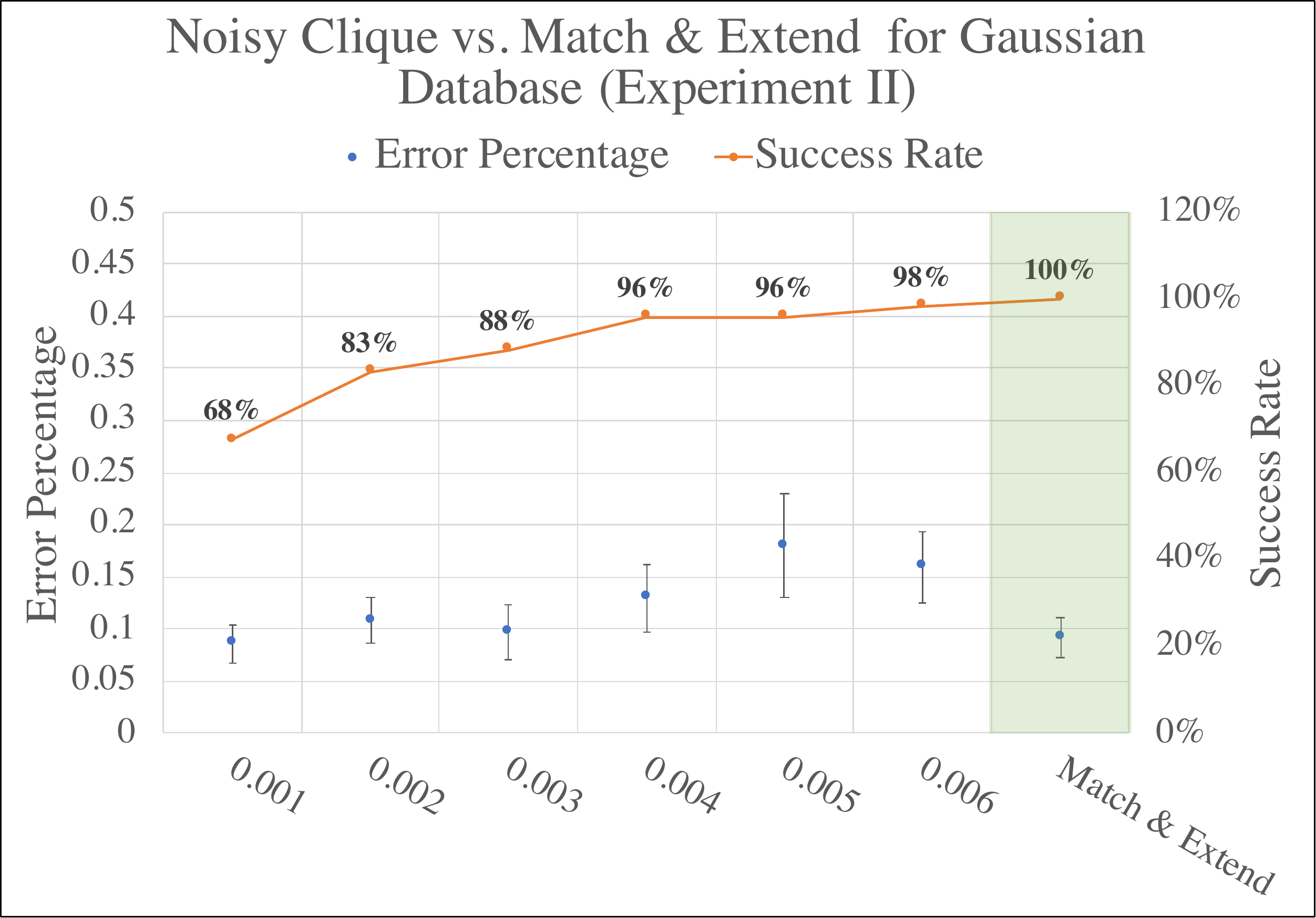}
\subcaption{Success rate and error percentage for Experiment II}
\label{fig:exp2}
\end{subfigure}
\hfill
\begin{subfigure}[t]{0.3\textwidth}
\includegraphics[width=\textwidth]{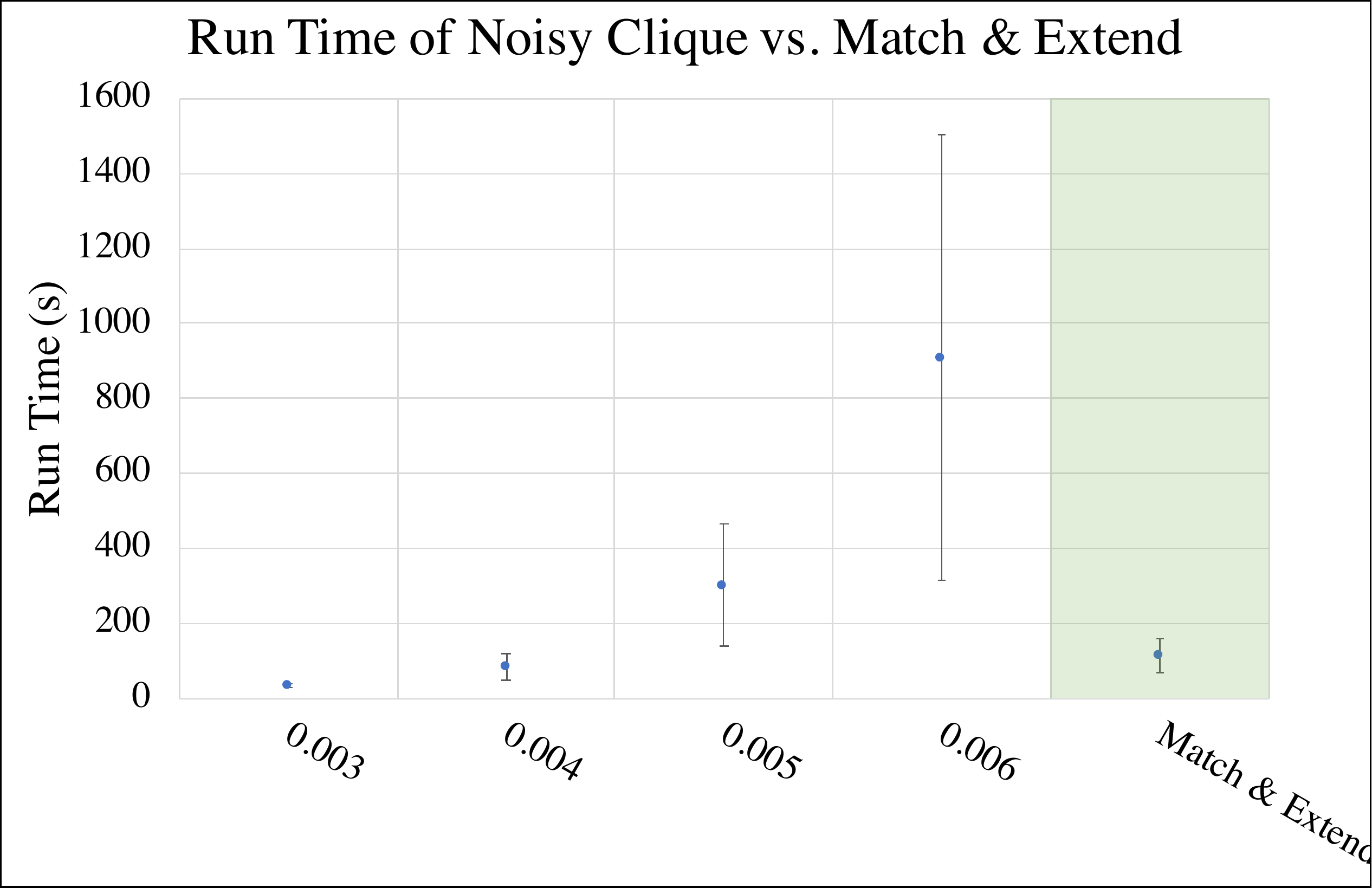}
\subcaption{Average running time of Noisy Clique for different Noise Budget vs. Match \& Extend Algorithm}
\label{fig:exp1-2_runtime}
\end{subfigure}

\caption{(a),(b) Comparison of the Success rate and Error Percentage for the noisy clique algorithm and Match \& Extend algorithm. The noisy clique algorithm is run using noise budget in the range $0.001-0.006$ and Match \& Extend is run with noise budget $0.002$. The orange line is the success rate (higher is better). The blue dot is the average error percentage and the black line segment is the $90\%$ confidence interval (lower dot and narrower interval is better).
(c) Average running time of Noisy Clique for different Noise Budget vs. Match \& Extend Algorithm with Noise Budget of $0.002$. 
\dnote{I don't understand what you mean by $90\%$ confidence?}
\mnote{this definition of the confidence interval was given before. I believe we can remove from here?}\anote{added in text}}
\label{fig:exp_all}
\end{figure*}

After processing the measurements, we obtained a set of approximate volumes, on which we then ran noisy clique-finding and Match \& Extend, which in turn output reconstructed databases. Figure~\ref{fig:exp1} illustrates the quality of the recovered values for the noisy clique-finding and Match \& Extend algorithms. 
The noisy clique-finding algorithm is run with several values for the noise budget while the Match \& Extend is run with a fixed noise budget of $0.002$. 
For each of the $N$ values $1, 2, \dots, N$, we expect to recover a candidate volume, corresponding to the number of records in the database that take that value. 
For a database with range of size $N$, we define the success rate as the number of candidate volumes recovered divided by $N$. 
For example in our experiments if we recover only $11$ candidate values for a database of range of size $N = 12$, then we have a success rate of $11/12$. It is also worth mentioning that the attacker can distinguish a successful attack from the failed one, since $N$, i.e. the size of the range, is known to the attacker. 
\mnote{changed the wording of Error Rate definition below}
We define the error rate of a recovered volume as its percentage of deviation from the original volume that it corresponds to.
We look at the recovered database and compare it to the original one. For each candidate volume $v'$ that is recovered, we compare it to the corresponding value in the real database, $v$ and report the error rate as $\left(\frac{|v'-v|}{v}\right) \times 100$.
\dnote{I don't understand the sentences above starting from ``For the Error Percentage.'' Tried to fix.}\anote{fixed}
So for example, if a recovered volume is $7990$ and the actual volume was $8000$ we report an error percentage of $0.12 \%$. 
If the algorithm only recovered $11$ values for a database of size $12$,
we will report the percentage error for the $11$ recovered values. 

Figure~\ref{fig:exp1} and~\ref{fig:exp2} show both the success rate and the error percentage for Experiment I and II. For \textit{success rate} (orange line), it can be seen that for the noisy clique-finding algorithm, increasing the noise budget helps to recover more volumes in both experiments.
The Match \& Extend algorithm, used with a fixed noise budget of $0.002$ could recover all the volumes in both of the experiments. 
\dnote{I don't understand the previous sentence.}\anote{tried to fix}
For \textit{error percentage} the average percentage of error is marked with a blue dot. The $90\%$ confidence interval is marked with the black marker. 
The confidence interval indicates that for a new set of experiments with the same setting, we are $90 \%$ confident that the average error rate will fall within that interval. \anote{tried to explain the $90 \%$ confidence interval and it needs better explanation} For the noisy clique-finding algorithm, increasing the noise budget causes the average error percentage to increase and the confidence interval to grow. 
\mnote{reworded the following sentence (original version commented)}
In some cases with noise budget $0.005$ and $0.006$, some of the recovered databases in Experiment I were very far off from the actual databases, causing the error interval in these settings to be much larger than in other settings.

In a nutshell, although it seems that increasing the noise budget helped to achieve higher success rates, since the error percentage grows, the quality of the recovered databases is lower. 
For the Match \& Extend algorithm the average amount of error and the width of error interval is comparable to the noisy clique-finding algorithm with small noise budget but the success rate is much higher. 
Figure~\ref{fig:exp1-2_runtime} shows the average run time as well as $90 \%$ confidence interval 
of the successful database recovery in seconds.
\dnote{I don't understand what you mean by $90\%$ confidence?}\anote{tried to fixed}
It can be seen that the average run time of noisy clique grows with the size of the noise budget. The Match \& Extend algorithm, however, always uses noise budget of $0.002$ and so its average running time remains low.

\begin{table}[]
\caption{Performance Comparison of Noisy Clique with Noise Budget $0.006$ vs. Match \& Extend Algorithm, with $99\%$ Confidence interval}
\footnotesize
\begin{tabular}{c|c|c|}
\cline{2-3} & \begin{tabular}[c]{@{}c@{}}Noisy Clique\\ (0.006)\end{tabular} & Match \& Extend \\ \hline
\multicolumn{1}{|c|}{Error Percentage} & 0.10 \% - 0.27 \% & 0.07 \% - 0.11 \% \\ \hline
\multicolumn{1}{|c|}{Run Time (s)} & 0 - 1900 & 38 - 190\\ \hline
\end{tabular}
\label{tab:perf}
\end{table}

Table~\ref{tab:perf} compares the performance of the Match \& Extend algorithm and the noisy clique algorithm on successful instances (meaning, the performance of the noisy-clique is taken only over the instances for which the recovered database had the correct size $N$). \dnote{I took this out, I don't think that's what you mean... ``with the same success rate.''}
The noisy clique-finding algorithm with noise budget $0.006$ performs better in terms of success rate than noisy clique-finding with smaller noise budgets, and we select it as a comparable algorithm to Match \& Extend algorithm. We are $99 \%$ certain that in a new set of experiments with the same setting as presented here, the Match \& Extend algorithm would output a result in at most $190$ seconds with at most $0.11 \%$ error. The noisy clique finding algorithm, on the other hand, would output a result in at most $1900$ seconds with at most $0.27 \%$ error. We have also analyzed the effect noise reduction step to further reduce the noise, which can be found in Section~\ref{sec:cvp_res}.

\subsection{Additional Noisy Process Running}\label{subsec:extraload}

\begin{figure}[t]
 \centering
  \includegraphics[width=0.8\linewidth]{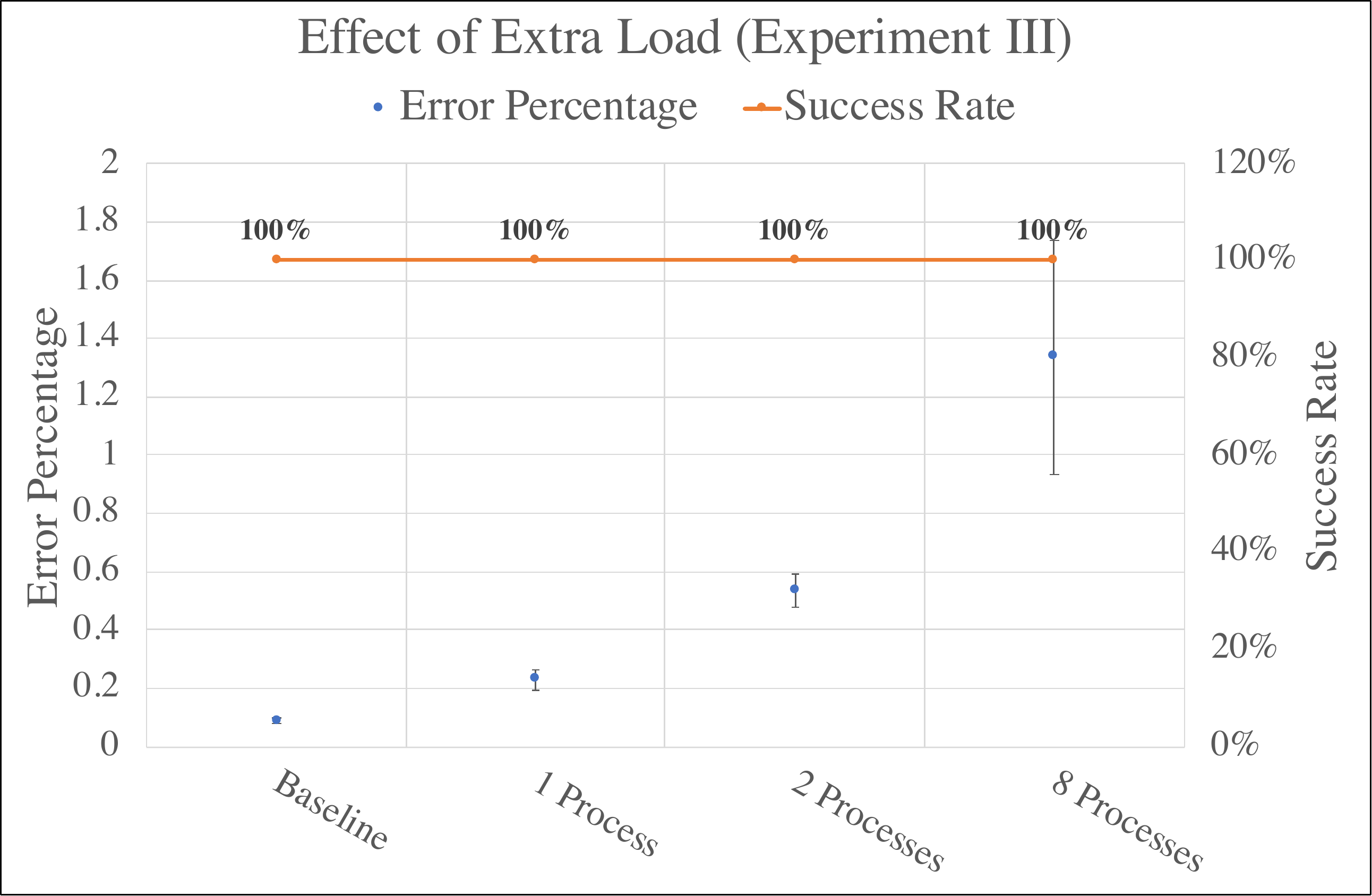}
  \caption{Success Rate and error percentage for Experiment III. The system is analyzed under different varying load. The load is increased by adding extra 1,2,8 processes. 
  }
  \label{fig:noisy_process}
\end{figure}

For Experiment III, we extend our analysis to the case where there is extra load on the system---i.e.~extra processes running---during the time of the attack. To run the noisy processes we use the command \texttt{stress -m i}, where $i$ represents the number of parallel threads running with CPU load of $100 \%$. We repeat the experiment for $i \in \{1,2,8\}$ and present the results in Figure~\ref{fig:noisy_process}. We took $5$ databases from Experiment I and obtained new sets of traces while the system had noisy processes running at the background. We recovered all the coordinates of the databases for all cases, so the success rate for all cases remains at $100 \%$. The quality of the recovered databases worsened with heavier loads, however even with $8$ noisy processes the average error remains less than $2 \%$. 

\begin{figure*}[htbp]
\centering
\begin{subfigure}[t]{0.30\textwidth}
\includegraphics[width=\textwidth]{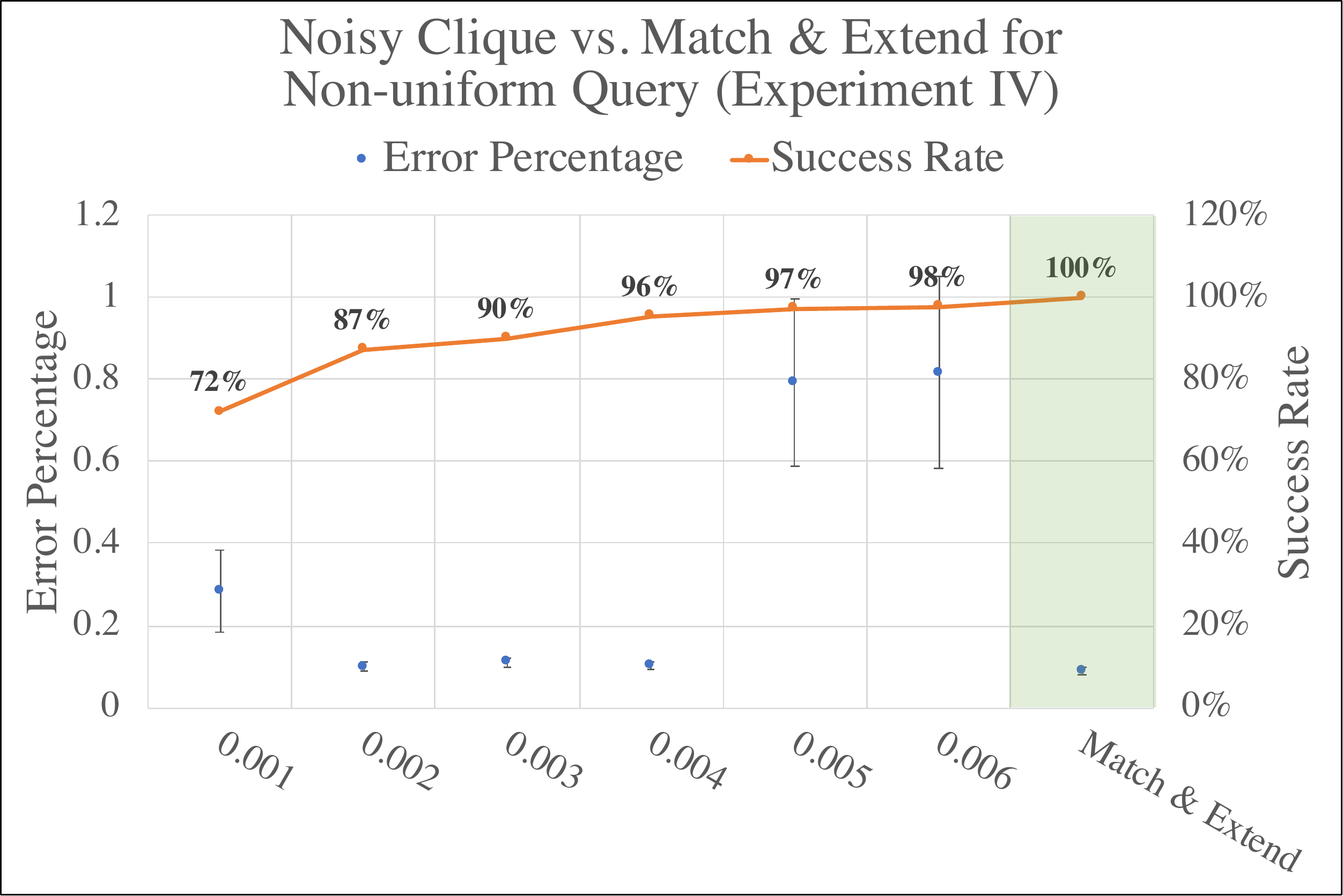}
\subcaption{Success Rate and error percentage for Experiment IV}
\label{fig:nonuni}
\end{subfigure}
\hfill
\begin{subfigure}[t]{0.30\textwidth}
\includegraphics[width=\textwidth]{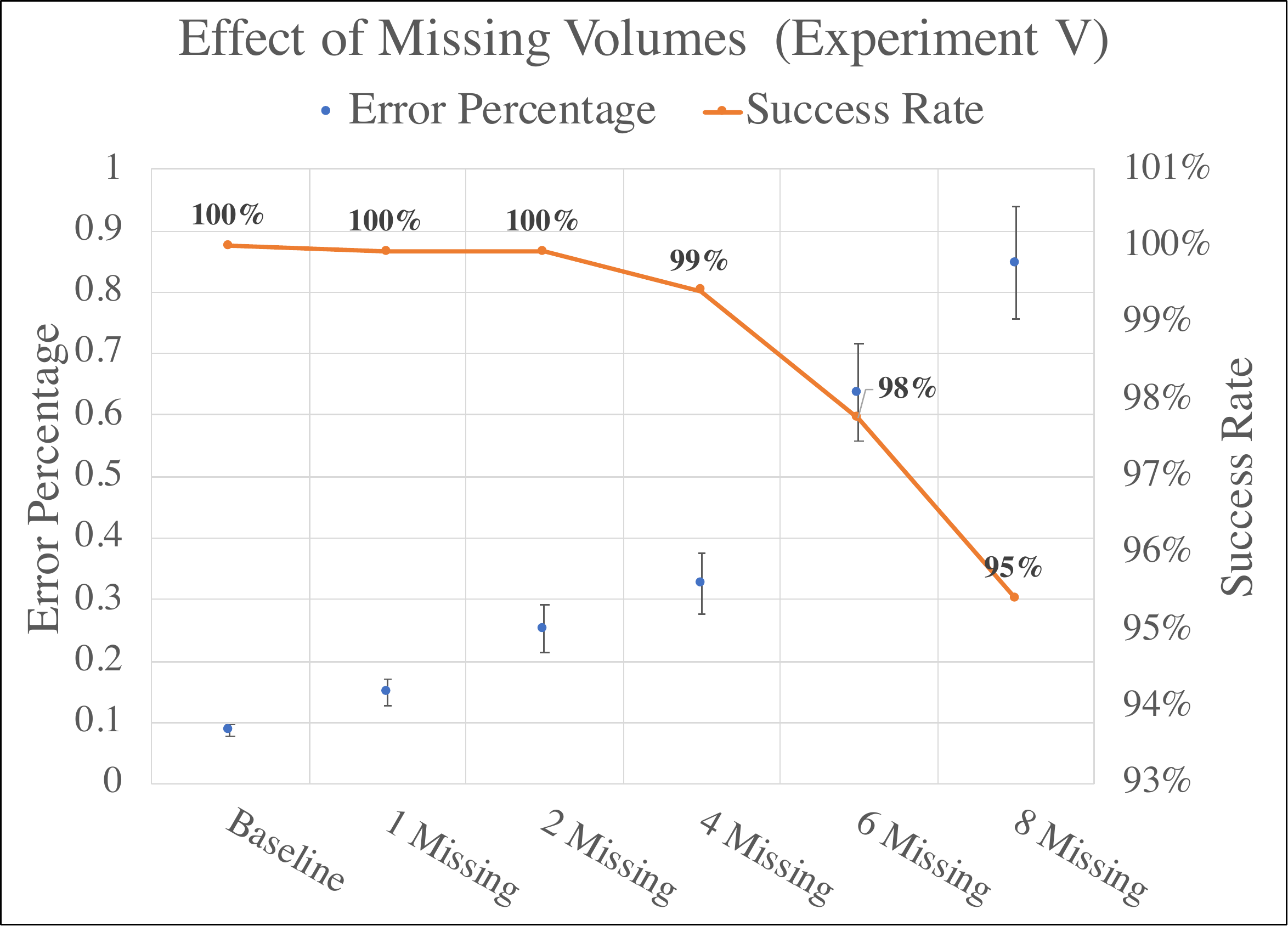}
\subcaption{Success rate and error percentage for Experiment V 
}
\label{fig:missingvols}
\end{subfigure}
\hfill
\begin{subfigure}[t]{0.30\textwidth}
\includegraphics[width=\textwidth]{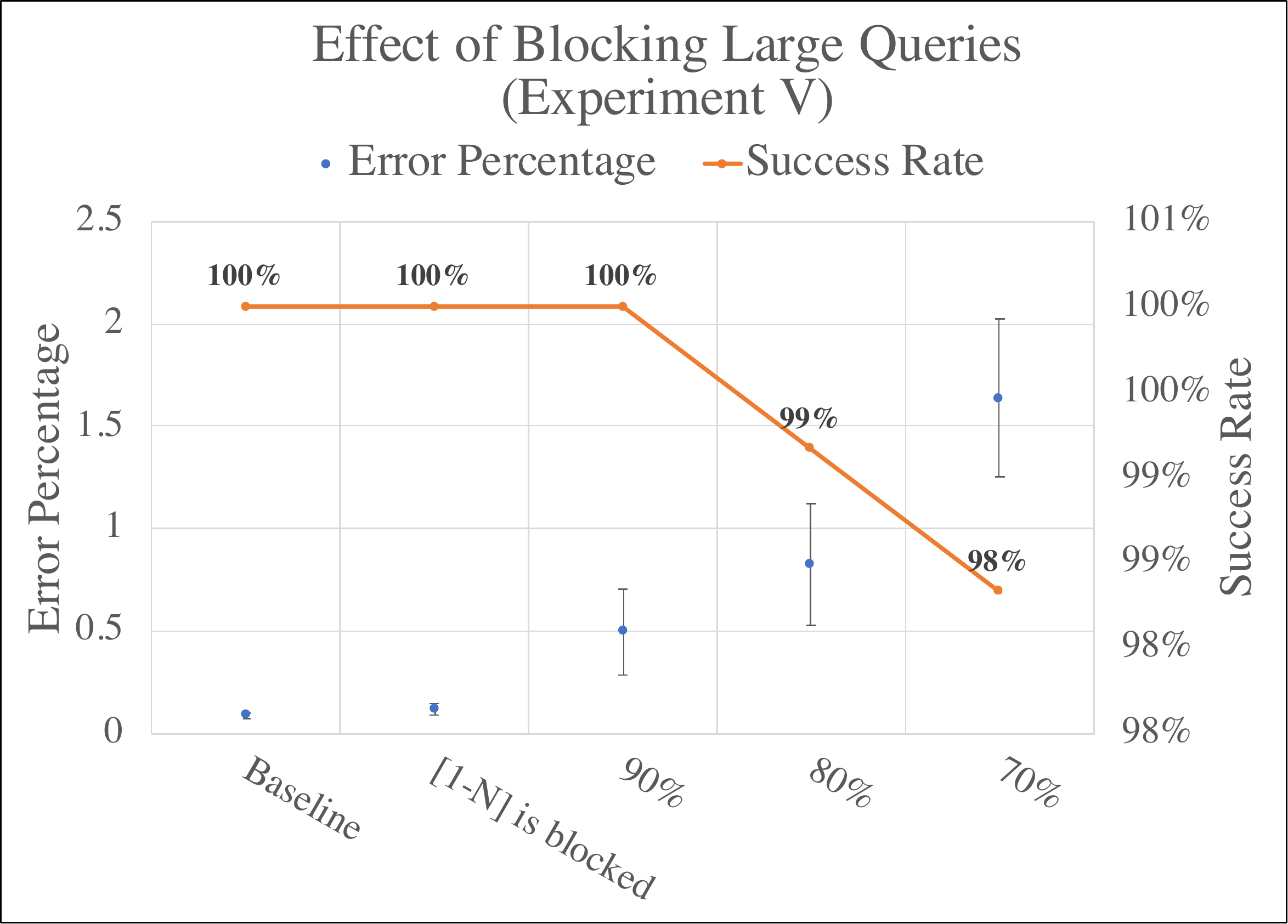}
\caption{Success rate and error percentage for Experiment V
}
\label{fig:missingbigvols}
\end{subfigure}

\caption{Performance of Match \& Extend algorithm in different cases (a) non-uniform query (b) Missing Volumes (c) Queries with large response size are blocked. The orange line is the success rate (higher is better). The blue dot is the average error percentage and the black line segment is the $90\%$ confidence interval (lower dot and narrower interval is better).
}
\label{fig:exp_all_noisy}
\end{figure*}

\subsection{Non-uniform Query Distribution}~\label{subsec:nonuni}
As mentioned, in order to be able to detect an approximate volume in our side-channel measurements, we need to observe at least $120$ measurements per range query. 
While we require that each query must be made some minimum number of 
times,
we do not impose that the query distribution must be uniform.
This is in contrast to previous work by Kellaris et al.~\cite{kellaris2016generic} who crucially required uniform query distribution. Our requirement can be seen as the noisy analogue of Grubbs et al.~\cite{grubbs2018pump}, who required each volume to be observed at least once in the \emph{noiseless} setting. To tolerate the noise present in the measurement in our setting, we require each query to be observed at least $120$ times. Later, in Subsection~\ref{subsec:missingvol}, we will further relax this requirement by showing that the recovery can be done even if some of the range queries are entirely omitted. This can be viewed as a weakening of even the requirements of Grubbs et al.~\cite{grubbs2018pump}.

For Experiment IV, we used the same databases as in Experiment I, but performed non-uniform range queries. We picked $5$ database from Experiment I and tested them with $3$ sets of non-uniform query distributions 
which results in $15$ scenarios in total. 
The first query distribution is chosen based on the assumption that queries of the form $\range{i}{i+1}$ are made twice as often as the other queries. The second query distribution assumed that queries in the form of $\range{i}{i+1}$ and $\range{i}{i+2}$ are made twice as often as the other queries. For the last query distribution, we tested the hypothesis that the determining factor for our attack seems to be the ability to identify ``peaks'' that roughly correspond to the volumes of the range queries (e.g.~see Figure~\ref{fig:volume_attack} and Figure~\ref{fig:aprox_vs_real}). A challenging query distribution is therefore one which causes one of the peaks to disappear as the peak adjacent to it dominates it. To test our hypothesis, we chose a distribution in which range $[a,b]$
was queried twice as often as 
$[c,d]$ when ranges $[a,b]$ and $[c,d]$ had
close volumes (and therefore close peaks). 

Figure~\ref{fig:nonuni} shows both the success rate and the error percentage for Experiment IV. It can be seen that the noisy clique-finding algorithm exhibits better success rate as the noise budget increases, while the error percentage grows as well. Match \& Extend algorithm, however has $100 \%$ success rate with low error percentage. We have also analyzed the effect of the noise reduction step to further reduce the noise, which can be found in Section~\ref{sec:cvp_res}. 


\subsection{Missing Queries}~\label{subsec:missingvol}

For Experiment V, we study the performance of our algorithm when some of the ranges are never queried. We consider two cases. In the first case, we look at a setting in which certain randomly chosen ranges
are never queried and in the second case, we consider a setting in which the queries corresponding to the largest volumes are never made. For the first case, we randomly drop $\{1,2,4,6,8\}$ volumes from the measurements. The performance of Match \& Extend algorithm can be seen in Figure~\ref{fig:missingvols}. As more and more volumes are dropped, the success rate of the algorithm decreases. For example, in the case where $8$ volumes are missing we recovered around $95 \%$ of the databases entries. However, the error percentage grows as more and more volumes are missing, although it remains below $1 \%$ error even for the case where $8$ volumes are missing. For the second case, we first look at the case where the query $\range{1}{N}$ is blocked, i.e.~the attacker is not allowed to query the whole database. In other cases the queries which ask for more than $90 \%, 80 \%$ and $70 \%$ of the database is blocked, respectively. It can be seen in Figure~\ref{fig:missingbigvols} that in these cases the success rate remains around $98 \%$ and the error percentage of the recovered coordinates stays below $2 \%$.

\section{Conclusions and Future Work}\label{sec:conclusion}

In this work we launched a cache side-channel attack against the SQLite database management system. We developed two algorithms that approximately recover the database using the information leaked from the side-channel attack. Finally, we showed the effectiveness of closest vector problem (CVP) solvers in reducing the overall noise in the recovered databases to obtain databases with improved accuracy. We showed that for attributes with range of size $12$ our algorithm can recover the approximate database in at most $190$ seconds with \dnote{average?} maximum error percentage of $0.11 \%$. We have also extended our analysis to study the effect of heavy load on the system as well as cases where some of the ranges are missing. We have shown that the error percentage for those cases remain below $2 \%$.
\dnote{The error percentage given here is contradicting what was given earlier.}\anote{I report the number from Table 3 now.}

As a possible approach to mitigate the attacks presented in this work, we suggest that when processing a range query, a random number of dummy elements get appended to the results and returned in addition to the true matches. The effect of such a countermeasure is twofold. (1) It makes it difficult for the side-channel attacker to able to aggregate information over different runs to obtain good approximations of the volumes. (2) It makes the graph generation and clique-finding algorithms more expensive, as there will be a large number of additional nodes and edges in the graph (recall that each observed volume corresponds to a node in the graph).
Since clique-finding is NP-hard, adding even a small fraction of nodes to the graph can make the attack infeasible.

As a future work, it would be interesting to explore the effectiveness of the attack using fewer traces; in the extreme case it is interesting to study the scenario where only $1$ trace per query is given. Moreover, it would be of interest to study the performance of the Prime \& Probe ~\cite{osvik2006cache} which is a more generic type of cache side-channel attack that can be used even in scenarios where the victim and attacker do not have a shared library. Further, as mentioned previously, improved attacks on encrypted databases are possible when the full access pattern is revealed (cf. Grubbs et al.~\cite{grubbs2019learning}). It will be interesting to explore whether partial information about the access pattern can be obtained via the cache side-channel and whether this information can be used to obtain improved attacks.
We have simulated some non-uniform query distribution, and one area that can be explored more is to study what are the most realistic query distribution.
One limitation we faced in the work is the scalability of solver for NP-hard problem, i.e. clique finding algorithm. The work of Grubbs et al.~\cite{grubbs2018pump} tolerate this by having a prepossessing step. This step enables the algorithm to work even for the cases where the size of the graph is large and it is interesting to study whether a prepossessing step can be applied to cases where volumes are noisy. 
\section*{Acknowledgments}
The authors would like to thank the members of the Dachman-Soled group in
REU-CAAR 2017
(funded by NSF grant \#CNS-1560193),
Stuart Nevans Locke,
Shir Maimon,
Robert Metzger,
Laura B.~Sullivan-Russett,
who helped us test
preliminary ideas
on cache side-channel attacks
for database reconstruction.
We also thank Uzi Vishkin
and Lambros Mertzanis for helpful discussions during various stages of this project. The authors would also like to thank the anonymous reviewers for their insightful comments and suggestions.




\bibliographystyle{plain}
\bibliography{main}

\appendix
\section{Appendix}\label{sec:appendix}

\paragraph{Closest Vector Problem (CVP)}
Given $n$-linearly independent vectors $\mathbf{b}_1, \mathbf{b}_2, \ldots, \mathbf{b}_n \in \mathbb{R}^m$, the lattice generated by $\mathbf{b}_1, \mathbf{b}_2, \ldots \mathbf{b}_n$ is the set of all the integer linear combination of them i.e. $\mathcal{L}(\mathbf{b}_1, \mathbf{b}_2, \ldots \mathbf{b}_n) = \{ \sum_{1}^{n}\mathbf{b}_i x_i\ | \ x_i \in \mathbb{Z}\}$. The set $\{ \mathbf{b}_1, \mathbf{b}_2, \ldots \mathbf{b}_n \}$ is called the basis of the lattice and is presented by matrix $\mathbf{B}$ in which basis $\mathbf{b}_i$ is $i$-th row of the matrix. In the closest vector problem target vector $\mathbf{y}$ is given. 
The target vector $\mathbf{y}$ does not necessarily belong to lattice $\mathcal{L}$. 
The solution is a lattice point $\mathbf{y}' = \mathbf{x}\mathbf{B}$ which is closest to target vector $\mathbf{y}$ and also $\mathbf{y}' \in \mathcal{L}$. Notice that a lattice point $\mathbf{y}'$ is a linear combination of basis, while the target vector $\mathbf{y}$ is not. The significance of the CVP problem is to find a closest vector to $\mathbf{y}$ such that the linear combination is satisfied.  CVP problem is also known to be NP-complete and we use \texttt{fplll}~\cite{fplll} for finding the closest vector in lattice.

\subsection{Error Reduction Step}\label{sec:CVP}

As explained in Section~\ref{sec:noisy_clique} and Section~\ref{sec:match_extend} by using the noisy clique-finding and Match \& Extend algorithms on the noisy data we get some \textit{close} answer to the real database. Here we outline a technique which can reduce the noise and output a \textit{more accurate} answer. The first step is to compute all the $\binom{N}{2} + N$ volumes corresponding to each range. Specifically, the ranges $\range{1}{1}, \range{1}{2}, \ldots, \range{1}{N}$ are obtained using noisy clique-finding or Match \& Extend. 
Each range $\range{i}{j}$ can be computed from the elementary volumes as $\size{\range{i}{j}} = \size{\range{1}{j}} - \size{\range{1}{i-1}}$. Instead of taking the computed value for range $\range{i}{j}$, we choose the value in the set of volumes (obtained from the side-channel data) that is closest to this computed value. This procedure results in $N' = \binom{N}{2} + N$ volumes which we call \emph{candidate volumes}.
Now note that given the volumes of the ranges $\range{1}{1}, \range{2}{2}, \ldots, \range{N}{N}$, the volume of any other range $\range{i}{j}$ can be expressed as a \emph{linear combination} of these values. Therefore, our variable $\vec{x} = (x_1, \ldots, x_N)$ corresponds to the volumes of the ranges $\range{1}{1}, \range{2}{2}, \ldots, \range{N}{N}$ and our candidate volumes $ \vec{v} = (v_1, \ldots, v_{N'})$, correspond to noisy linear combinations of the $x_i$'s. Thus, solving for the $\vec{x}$ which yields the closest solution to $ \vec{v} = (v_1, \ldots, v_{N'})$
under the linear constraints, corresponds to solving a Closest Vector Problem (CVP).

For example, if the range has size $N = 3$, then we obtain a total of $6$ volumes $v_1, \ldots, v_6$ corresponding to the ranges $\range{1}{1}$, $\range{2}{2}$, $\range{3}{3}$, $\range{1}{2}$, $\range{2}{3}$, $\range{1}{3}$ and can construct the following system of equations:
\[
\mathbf{A} \vec{x} + \vec{e} = \vec{v}
\]

where 
\[
\mathbf{A}=
\left[ {\begin{array}{ccc}
1 &0 &0\\
0 &1 &0\\
0 &0 &1\\
1 &1 &0\\
0 &1 &1\\
1 &1 &1
\end{array} } \right]
\]
$\vec{v} = (v_1, \ldots, v_6)$, $\vec{e}$ is the amount of error and $\vec{x}$ is unknown. To solve this problem, we can consider the lattice defined by $\mathbf{A}\vec{z}$, where $\mathbf{A}$ is the basis and $\vec{z}$ is any integer vector. Now, given $\vec{v}$, we would like to find the closest lattice vector $\vec{y} = \mathbf{A}\vec{x'}$ to $\vec{v}$. Once we have $\vec{y}$, we can solve to get $\vec{x'}$.
To create a full rank matrix for our solver, we can modify matrix $A$ as following:
\[
A' =
\left[ {\begin{array}{cccccc}
1 &0 &0 &0 &0 &0\\
0 &1 &0 &0 &0 &0\\
0 &0 &1 &0 &0 &0\\
1 &1 &0 &T &0 &0\\
0 &1 &1 &0 &T &0\\
1 &1 &1 &0 &0 &T
\end{array} } \right]
\]
where $T \gg n$ and $\vec{v}$ stays the same.

Now we obtain a solution of dimension $6$ (as opposed to dimension $3$), but the last three coordinates should always be $0$, since if they are non-zero there will be at least $\pm$ $T$ in the corresponding coordinate of $\vec{v}$ \mnote{did we mean coordinate of $\vec{y}$?}\anote{it should be okay...}\mnote{can we review this part again? Maybe above we could say $\vec{v'}=\mathbf{A}\vec{x'}$, and here say "in the corresponding coordinate of $\vec{v'}$}, which will clearly not be the closest vector.

\subsection{Error Reduction Step Experiments}~\label{sec:cvp_res}

Table~\ref{tab:l1perf} and Table~\ref{tab:linfperf} compare the clique and Match \& Extend algorithms and the improvement achieved by the error reduction step using the CVP solver.
Recall that the error percentage is computed for each recovered coordinate.
\mnote{reworded the following explanation of Tables 2 and 3 in hope to make it more understandable (previous version commented out)}
We measured the quality of the recovered databases in two ways: in Table~\ref{tab:l1perf} we report the average value of the error percentage over all the volumes in all recovered databases. For Table~\ref{tab:linfperf}, we compute for each database the largest error percentage of its coordinates, and we report the average of all these maxima over all databases in Experiments I, II and IV. The effectiveness of CVP is for the cases where the initial reconstructed database is within some \emph{close} distance of the correct database. Hence for the other two experiments, the plain CVP is not effective as the initial recovered database is rather far from the correct answer, and generally CVP can not be effective. Moreover, the CVP algorithm needs to have all the initial volumes, so for the experiments where some of the volumes are missing, CVP can not be used.

It can be seen that for Match \& Extend algorithm the average error percentage is reduced from $0.09$ to $0.07$, from $0.09$ to $0.08$ and from $0.09$ to $0.08$ in Experiments I, II and IV, respectively. 
Table~\ref{tab:linfperf} shows similar results for maximum error percentage. Namely for Match \& Extend algorithm the maximum error percentage is reduced from $0.22$ to $0.20$, from $0.30$ to $0.22$ and from $0.24$ to $0.22$ in Experiments I, II and IV, respectively.\footnote{Table~\ref{tab:l1perf} and Table~\ref{tab:linfperf} present the $L_1$ norm and $L_{\infty}$ norm, respectively. The CVP solver optimizes for $L_2$ norm, so it has a larger effect on decreasing $L_{\infty}$ norm than $L_1$ norm. In case the objective is to minimize the $L_1$ norm, an integer programming approach would be preferable.}
\dnote{I think $L_\infty$ norm will be confusing for some readers. Should we just say the maximum deviation or something like that?}\anote{changed notation}

\dnote{We should also mention somewhere that the CVP solver that we use is for $L_2$ norm and so we expect to see a larger improvement in $L-\infty$ norm versus $L_1$ norm. Maybe say something about running an integer program for optimizing $L_1$ norm}\anote{added a footnote}

\dnote{For Tables 2 and 3, need to explain how the percentages are calculated. I.e. what are you dividing by? I also put in percent sign to emphasize.}\anote{added}

\begin{table}[]
\caption{Average Error Percentage}
\footnotesize
\begin{tabular}{l|c|c|c|c|}
\cline{2-5}
 & \multicolumn{2}{c|}{Noisy Clique}  & \multicolumn{2}{c|}{Match \& Extend} \\ \hline
\multicolumn{1}{|l|}{Experiment} & No CVP & CVP & No CVP & CVP \\ \hline
\multicolumn{1}{|l|}{Experiment I} & 0.21\% & 0.19\% & 0.09\% & 0.07\%\\ \hline
\multicolumn{1}{|l|}{Experiment II} & 0.16\% & 0.10\% & 0.09\% & 0.08\%\\ \hline
\multicolumn{1}{|l|}{Experiment IV} & 0.80\% & 0.77\% & 0.09\% & 0.08\%\\ \hline
\end{tabular}
\label{tab:l1perf}
\end{table}

\begin{table}[]
\caption{Maximum Error Percentage}
\footnotesize
\begin{tabular}{l|c|c|c|c|}
\cline{2-5}
 & \multicolumn{2}{c|}{Noisy Clique}  & \multicolumn{2}{c|}{Match \& Extend} \\ \hline
\multicolumn{1}{|l|}{Experiment} & No CVP & CVP & No CVP & CVP \\ \hline
\multicolumn{1}{|l|}{Experiment I} & 0.86\% & 0.86\% & 0.22\% & 0.20\% \\ \hline
\multicolumn{1}{|l|}{Experiment II} & 0.55\% & 0.38\% & 0.30\% & 0.22\%\\ \hline
\multicolumn{1}{|l|}{Experiment IV} & 2.29\% & 2.30\% & 0.24\% & 0.22\%\\ \hline
\end{tabular}
\label{tab:linfperf}
\end{table}

\end{document}